\documentclass[superscriptaddress,longbibliography,nofootinbib,amsmath,amssymb,prb,reprint,aps,floatfix]{revtex4-2}

\usepackage{graphicx}
\usepackage{xcolor}
\definecolor{newBlue}{HTML}{0072B2}
\definecolor{newMaroon}{HTML}{882255}
\usepackage[colorlinks,linkcolor=newBlue,urlcolor=newBlue,citecolor=newMaroon]{hyperref}

\begin{document}

\title{Fractional quantum spin Hall crystals from hidden density-wave order}

\author{Matthew Shammami}
\affiliation{Department of Chemistry \& Biochemistry,
University of California, Los Angeles, California 90095-1569}
\affiliation{Mani L. Bhaumik Institute for Theoretical Physics, University of California Los Angeles, Los Angeles, California 90095-1547}
\author{Sudip Chakravarty}
\affiliation{Mani L. Bhaumik Institute for Theoretical Physics, University of California Los Angeles, Los Angeles, California 90095-1547}
\affiliation{Department of Physics \& Astronomy, University of California Los Angeles, Los Angeles, California 90095-1547}

\date{\today}

\begin{abstract}
Broken symmetry and topology are distinct descriptions for how matter organizes itself. Here, we propose one gives rise to the other. A mixed singlet--triplet $d$-density wave, which is a particle--hole condensate, has a hidden bond order that breaks translation symmetry and generates a time-reversed pair of Chern bands, without producing conventional charge or spin density order. At fractional fillings with odd denominators, finite size exact diagonalization studies show how repulsive interactions can transform this symmetry-broken topological state into a fractional quantum spin Hall liquid. Its opposite chiralities produce a crossed response in which charge flux transports spin, and spin flux transports charge. The same setting also reveals what borders such a phase. Strong interspin coupling and increasingly dilute fillings favor charge order and make residual band dispersion decisive, while at half filling, the spin-decoupled limit gives a homogeneous correlated liquid with signatures consistent with the physics of a composite Fermi liquid. 
\end{abstract}

\maketitle

\section{Introduction}
 Broken symmetry and topology are core principles for organizing phases of matter. In a symmetry-broken phase, the system selects one of several states related by a symmetry of the underlying Hamiltonian, and the order is encoded in quantities --whether local or global -- that transform nontrivially under that symmetry. A topological phase is instead distinguished by global properties of the many-body wavefunction and is characterized by invariance with respect to smooth deformations of the Hamiltonian. Though we are so accustomed to treating these as two different paradigms in the theory of condensed matter, they need not be mutually exclusive. The quantum Hall effect provides the canonical example of this: Landau levels are generated by applying an external, symmetry-breaking field and carry nonzero Chern numbers. When filled, they exhibit quantized Hall conductance \cite{Klitzing1980,Laughlin1981,Thouless1982,Avron1983}, while interactions within partially filled levels may produce fractionalized phases \cite{Tsui1982,Laughlin1983, Laughlin1999}. The integer quantum \emph{spin} Hall (IQSH) effect showed that much of this physics can survive without the need to break time-reversal symmetry \cite{KaneMele2005a,KaneMele2005b,Bernevig2006}. In its early considerations, however, the topology of the problem was either imposed externally or engineered into noninteracting band structures. This raises a different question: can the \emph{spontaneous} breaking of symmetry itself generate topological bands which may then support integral or fractional quantum Hall responses?


Several mechanisms have been proposed for interaction-driven topological phases \cite{Raghu2008,GroverSenthil2008,Sun2009,Zhang2009, Yang2010}. Here, we revisit a case where topology is inseparable from the broken symmetry order parameter itself, introduced by one of us and co-workers in Ref.~\cite{Hsu2011}. The state is a $d$-density wave of mixed singlet and triplet spin character, which is a particle-hole condensate of nonzero ($\ell=2$) angular momentum in the continuum classification \cite{Nayak2000}. Although the state breaks both translation and spin rotational symmetries, its form factors produce neither charge nor spin density modulations. The order lives instead in staggered spin currents \cite{Nersesyan1991} and modulations of diagonal bond kinetic energy, making it a hidden order since conventional $s$-wave probes do not couple linearly to its order parameter. The coexistence of the imaginary spin triplet $d_{x^2-y^2}$ and real singlet
$d_{xy}$ components gaps the density-wave reconstructed spectrum and gives rise to a time-reversed pair of Chern bands, with spin Chern numbers $(C_{\uparrow}, C_{\downarrow})=(-1, 1)$. At $\nu=1$, corresponding to having one electron per lattice site, the lower spin-degenerate pair is filled (i.e., half filling of the original spinful lattice). The charge Hall responses of the two spin sectors cancel, while their spin Hall responses add, realizing the IQSH effect with quantized spin Hall conductance $|\sigma_{xy}^{\mathrm{spin}}|=e/2\pi$.


The natural question is whether these ideas can be extended to fractional values of $\nu$; can these bands, when partially filled, realize fractional
quantum spin Hall (FQSH) states? FQSH phases were  proposed using trial wavefunctions and effective topological field theories \cite{Bernevig2006,LevinStern2009}, and later investigated numerically in interacting lattice models \cite{Neupert2011,Wei2014,Repellin2014}. This would be an idle curiosity, although exceedingly interesting, if it were not for tantalizing experimental signatures of such states at zero field in moir\'e twisted bilayer MoTe$_2$ \cite{Kang2024, Wang2026}. We do not attempt a microscopic description of a specific experimental setting. Rather, we ask whether residual interactions can fractionalize a time-reversed pair of Chern bands that are themselves generated by a broken symmetry condensate. This scenario is broadly similar in spirit to the anomalous Hall crystal descriptions, in which spontaneous breaking of translation and time-reversal symmetries generates the parent Chern band, whose partial filling may explain the charge fractionalization observed in rhombohedral multilayer graphene \cite{Lu2024,Dong2024,DongPatriSenthil2024, ZhouYangZhang2024}. Historically, the coexistence of broken translation symmetry and quantized Hall response in a magnetic field was established by Halperin and co-workers \cite{Halperin1986,Tesanovic1989}; see also recent work on this coexistence at integer and fractional fillings in Ref.~\cite{maymann2026}.


We take the singlet-triplet $d$-density wave (stDDW) condensate as a parent order and select a single uniform domain, leaving its self-consistent formation and collective fluctuations outside the scope of the current analysis. This isolates the question of what happens when residual interactions act within its partially filled Chern bands. We project short-range density interactions into the lower band pair and study the resulting many-body problem by exact diagonalization on finite tori at fixed $S_z$. Momentum-resolved spectra and spin-independent and spin-dependent flux threading are combined with density correlations to distinguish fractional liquids from competing ordered states. At $\nu=1/3$, same-spin repulsion produces the expected decoupled FQSH manifold even in a dispersive band. Comparable opposite-spin repulsion replaces it with an isolated triplet, but the accompanying charge correlations favor a charge order interpretation rather than a homogeneous, coupled FQSH liquid. At $\nu=1/5$, the Laughlin-like structure survives in the flat-band limit but is destabilized by electron dispersion, while no corresponding fractional manifold appears on the numerically accessible $\nu=1/7$ clusters. At $\nu=1/2$, same-spin repulsion instead strongly fragments the band-projected spectral weight and distributes it among many excitations, without producing density order on the computationally accessible clusters. The state is likely a compressible composite Fermi liquid in each spin sector, with the two sectors related by time-reversal symmetry, as opposed to a weakly renormalized electron metal. Upon adding strong interspin repulsions, the system eventually
crosses into fully polarized
quantum anomalous Hall sectors.


The plan of the paper is as follows. In \hyperref[sec:tdw]{Sec.~II} we introduce the mixed singlet--triplet density-wave state and its integer quantum spin Hall response. In \hyperref[sec:fractionalization]{Sec.~III} we project density-density interactions into the lower stDDW bands and present the exact diagonalization results for different fillings. \hyperref[sec:conclusion]{Sec.~V} summarizes the results and discusses remaining questions. There are three appendices.

\section{The topological density wave state}
\label{sec:tdw}

Density-wave order can be understood as condensation in the particle-hole channel, an idea developed for excitonic insulators by Halperin and Rice and later generalized for lattice condensates of nonzero angular momentum by Nayak \cite{Halperin1968,Nayak2000}. In an order parameter $\langle c^\dagger_{\mathbf{k}+\mathbf Q,\alpha}c_{\mathbf{k},\beta}\rangle$, the wavevector $\mathbf Q$ determines how the condensate transforms under lattice translational symmetry, while the dependence on $\mathbf{k}$ describes the internal form factor of the particle-hole pair. A momentum-independent form factor gives the usual charge- or spin-density waves, whereas a nontrivial lattice harmonic (such as the $d$-wave form factors considered here, which have no on-site component) places the order on bonds, as modulations of kinetic energy or bond currents. Unlike a Cooper pair, the spin and orbital makeup of a particle-hole bilinear need not obey an exchange-antisymmetry constraint, so a $d$-wave particle-hole condensate may be either spin singlet or spin triplet, allowing the coexistence of the hybrid state considered here.


On the square lattice, the commensurate imaginary spin-singlet $d_{x^2-y^2}$ condensate is the orbital antiferromagnet, or $d$-density wave (DDW), proposed as a hidden-order description of the cuprate pseudogap and studied in interacting lattice models \cite{Chakravarty2001,NayakPivovarov2002,Laughlin2014}. Its spin-triplet counterpart is a fermionic spin nematic state whose order lives in staggered spin currents \cite{Nersesyan1991}. Ref.~\cite{Hsu2011} combined this triplet component with a real singlet $d_{xy}$ bond order to obtain the gapped topological density wave studied below.

\subsection{The mixed singlet and triplet density wave}

We set the square lattice spacing to unity and take the commensurate ordering wavevector to be $\mathbf Q=(\pi,\pi)$. A nonzero particle-hole condensate $\langle c^\dagger_{\mathbf{k}+\mathbf Q,\sigma}c_{\mathbf{k},\sigma'}\rangle$ mixes states at $\mathbf{k}$ and $\mathbf{k}+\mathbf Q$. Following Ref.~\cite{Hsu2011}, the associated mean-field gap matrix in the reduced Brillouin zone (RBZ) can be written as
\begin{align}
D_{\sigma\sigma'}(\mathbf{k})
&=
\delta_{\sigma\sigma'}
\big(\Delta_{\mathbf{k}}+i\eta_\sigma W_{\mathbf{k}}\big),
\end{align}
where
\begin{align}
\Delta_{\mathbf{k}}
&=
\Delta_0\sin k_x\sin k_y,
\qquad
W_{\mathbf{k}}
=
\frac{W_0}{2}(\cos k_x-\cos k_y),
\label{eq:form_factors}
\end{align}
with $\Delta_0,W_0$ being real, energy-valued amplitudes and $\eta_\sigma=\pm 1$ for $\sigma=\uparrow,\downarrow$, respectively. Note that here we have chosen a uniform triplet orientation $\hat{\mathbf n}=\hat{\mathbf z}$, since more generally, the triplet component is $iW_{\mathbf{k}}\hat{\mathbf n}\cdot\mathbf s$ (with $\mathbf{s}$ being the Pauli matrices in spin space). For this rotation-ivariant Hamiltonian, choosing $\hat{\mathbf n}$ breaks $SU(2)$ down to $U(1)$, giving the order parameter manifold $SU(2)/U(1)\simeq S^2$. Here we have chosen one such domain, $\hat{\mathbf n}=\hat{\mathbf z}$ with conserved $U(1)_{S_z}$ and neglected fluctuations of $\hat{\mathbf n}$ in our mean-field ansatz.

\begin{figure}[t]
\centering
\includegraphics[width=0.95\linewidth]{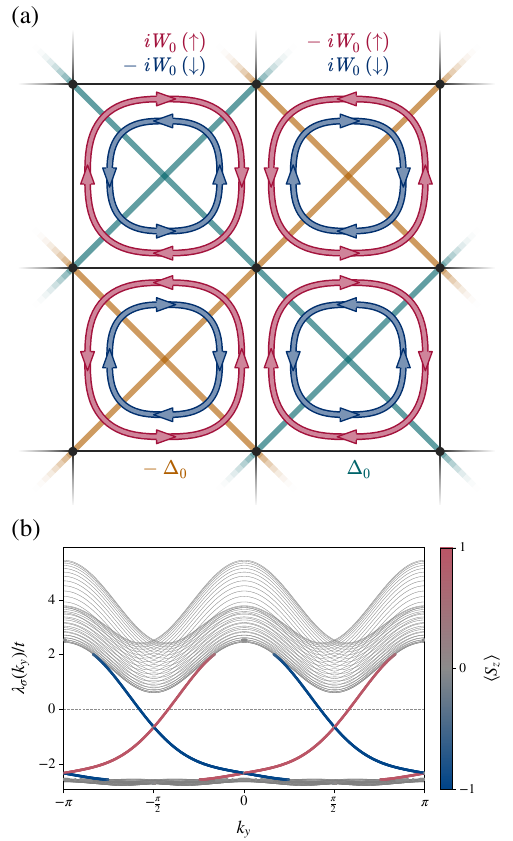}
\caption{(a) The mixed singlet-triplet $d$-density wave model in real space. The triplet component produces opposite staggered circulations for the two spins, whereas the singlet one modulates the diagonal bonds. (b) Edge state dispersion, computed from a ribbon geometry that is 40 sites wide and periodic along $y$. The spectrum shows a pair of helical edge states crossing the bulk gap per edge, with the color denoting their edge-resolved spin polarization.}
\label{fig:lattice}
\end{figure}

\begin{figure}[t]
\centering
\includegraphics[width=\linewidth]{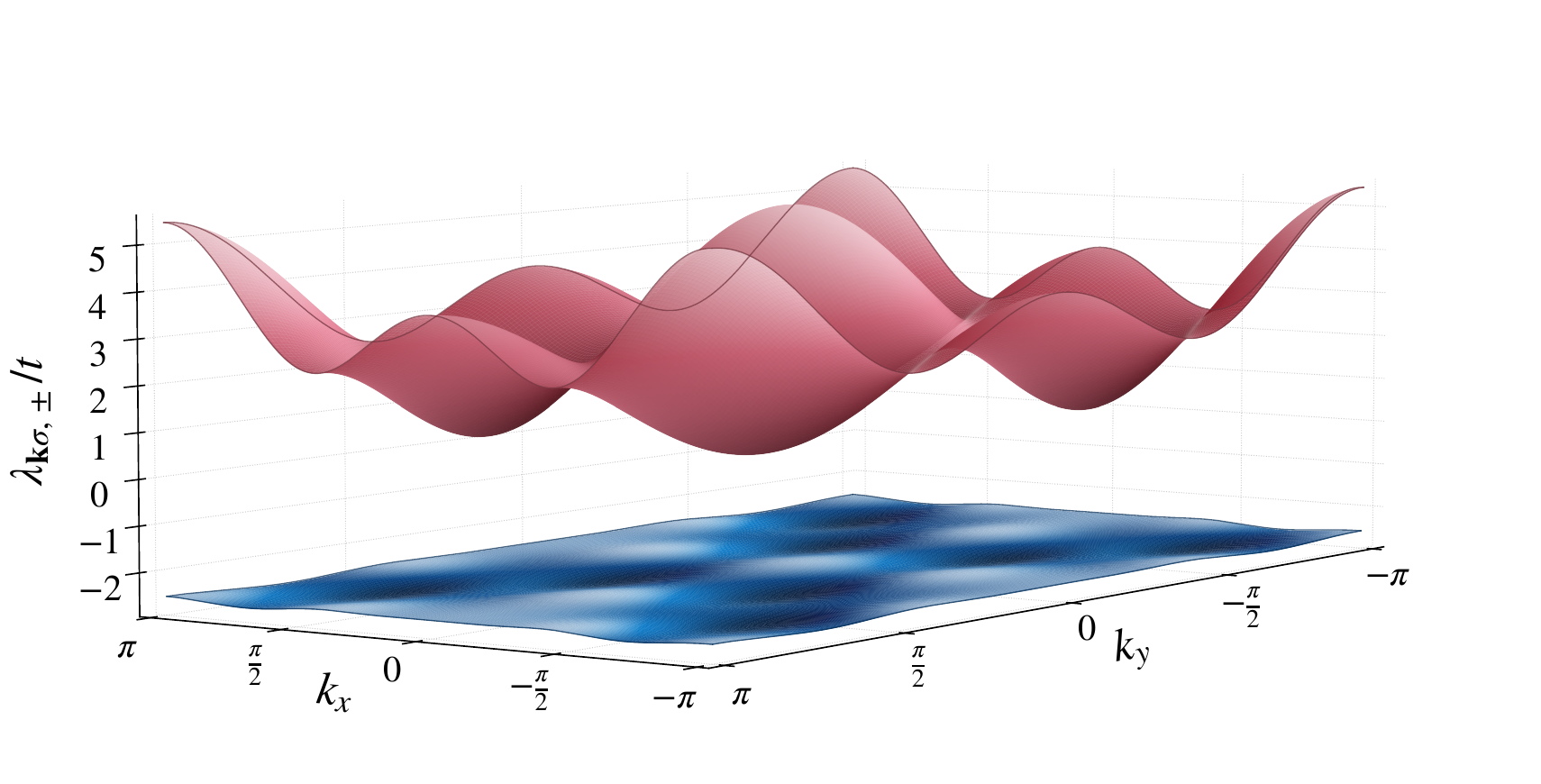}
\caption{Bulk band structure of the mixed singlet--triplet density-wave model for the parameters listed in Sec.~\ref{sec:nu_1_3}.}
\label{fig:band}
\end{figure}

We supplement this density-wave order with nearest-neighbor, next-nearest-neighbor, and axial third-neighbor hoppings $t$, $t'$, and $t''$. The dispersion splits into parts that are odd and even under $\mathbf{k}\mapsto\mathbf{k}+\mathbf Q$: $\varepsilon_{1\mathbf{k}}=-2t(\cos k_x+\cos k_y)$ and $\varepsilon_{2\mathbf{k}}=4t'\cos k_x\cos k_y+2t''(\cos 2k_x+\cos 2k_y)$. Restricting $\mathbf{k}$ to the RBZ and defining $\psi^\dagger_{\mathbf{k}\sigma}=(c^\dagger_{\mathbf{k}\sigma},c^\dagger_{\mathbf{k}+\mathbf Q,\sigma})$, the Hamiltonian and eigenvalues for each spin sector may then be given by:
\begin{align}
h_\sigma(\mathbf{k})
&=
(\varepsilon_{2\mathbf{k}}-\mu)\tau_0
+\varepsilon_{1\mathbf{k}}\tau_3
+\Delta_{\mathbf{k}}\tau_1
-\eta_\sigma W_{\mathbf{k}}\tau_2,\nonumber\\
\lambda_{\mathbf{k}\sigma,\pm}
&=
\varepsilon_{2\mathbf {k}}-\mu
\pm E_\mathbf{k}
\label{eq:H_MF}
\end{align}
where $E_\mathbf{k}=\sqrt{\varepsilon_{1\mathbf{k}}^2+\Delta_{\mathbf{k}}^2+W_{\mathbf{k}}^2}$.
Here $\boldsymbol{\tau}$ are the Pauli matrices acting on the $(\mathbf{k},\mathbf{k}+\mathbf Q)$ pseudospin. The spectrum is doubly degenerate throughout the Brillouin zone by $PT$ symmetry, and for nonzero $t, W_0$, and $\Delta_0$, the bands are separated by a direct gap of $2E_\mathbf{k}$, as shown in Fig.~\ref{fig:band}.

How this mixed order parameter gives rise to topology can be understood by looking at the effect of the singlet component in gapping out the Dirac points that are otherwise present in the semimetallic, triplet case. When $\Delta_0=0$, the direct gap closes at $(\pm\pi/2,\pm\pi/2)$ in the full Brillouin zone, corresponding to two inequivalent Dirac points in the RBZ. The $d_{xy}$ component takes opposite values or ``masses'', $\Delta_0$ and $-\Delta_0$, at these two points. Because the Dirac cones also have opposite chiralities, their contributions to the Chern number add. A momentum-independent singlet mass would instead have the same sign at both cones and give cancelling contributions. For the occupied lower bands,
$C_\sigma=\eta_\sigma\operatorname{sgn}(tW_0\Delta_0)$.
Time reversal exchanges the two spin blocks, $h_\downarrow(\mathbf{k})=h_\uparrow^*(-\mathbf{k})$, so the charge Chern number vanishes while the spin Hall conductance has magnitude $e/(2\pi)$ (see Appendix \ref{app:hall} for details). The dispersion of the edge states in this model contains one Kramers pair of helical edge modes [Fig.~\ref{fig:lattice}(b)]. Thus, $W_0$ gives a spin-dependent Dirac chirality, while $\Delta_0$ provides the mass that converts the reconstructed semimetal into a quantum spin Hall insulator.


One can also understand this order in real space. The density-wave part of Eq.~\eqref{eq:H_MF} can be written as $H_{\mathrm{DW}}
=\sum_{\mathbf{k}\in\mathrm{BZ},\sigma}
\bigl(\Delta_{\mathbf{k}}-i\eta_\sigma W_{\mathbf{k}}\bigr)
c^\dagger_{\mathbf{k}+\mathbf Q,\sigma}c_{\mathbf{k}\sigma}$.
Let $t^{(\sigma)}_{\mathbf r\rightarrow\mathbf r+\boldsymbol\delta}$ denote the coefficients of $c^\dagger_{\mathbf r+\boldsymbol\delta,\sigma}c_{\mathbf r\sigma}$ and write $s_{\mathbf r}=(-1)^{r_x+r_y}$. Fourier transforming, the hopping amplitudes are modified according to 
\begin{align}
t^{(\sigma)}_{\mathbf r\rightarrow\mathbf r+\hat{\mathbf x}}
&=
-t+i\eta_\sigma s_{\mathbf r}\frac{W_0}{4},
&
t^{(\sigma)}_{\mathbf r\rightarrow\mathbf r+\hat{\mathbf x}+\hat{\mathbf y}}
&=t'-s_{\mathbf r}\frac{\Delta_0}{4}
\end{align}
and the opposite signs for the $W_0$ and $\Delta_0$ terms for the $\hat{\mathbf y}$ bond and $\hat{\mathbf x}-\hat{\mathbf y}$ diagonal, respectively. Going around an elementary plaquette counterclockwise from $\mathbf{r}$, one accumulates a gauge invariant phase of
$-4\eta_\sigma s_{\mathbf r}\arctan(W_0/4t)$ (mod $2\pi$), which flips between neighboring plaquettes and between opposite spins. The charge currents cancel by time-reversal, giving the staggered spin currents shown in Fig. \ref{fig:lattice}(a). The imaginary spin-dependent hopping, which is generated spontaneously by the translation-symmetry breaking particle-hole condensate, therefore, plays a role reminiscent of the microscopic spin-orbit coupling term introduced by the Kane--Mele model.

\section{Interaction-driven fractionalization}
\label{sec:fractionalization}

We now partially fill the lower two bands and ask whether the interactions within them can stabilize incompressible fractional states. Let $N=N_xN_y$ be the number of single-particle states in each spin-resolved lower band. The wavevector $\mathbf Q=(\pi,\pi)$ doubles the unit cell, and the microscopic lattice has $N_s=2N$ sites. The total number of electrons ($N_e$) in the spinful Hilbert space is $N_\uparrow+N_\downarrow$. The total filling of the band pair is defined as $\nu={N_e}/{2N}$ (or $\nu_\sigma=N_\sigma/N$ per component). For example, a $3\times4$ RBZ grid has $N=12$ orbitals per band and $N_{\mathrm s}=24$ microscopic sites. At $\nu=1/3$, it contains $N_\uparrow=N_\downarrow=4$ and $N_e=8$ in an $S_z=0$ sector.


We take the forces governing electron interactions to be simply nearest-neighbor repulsions,
\begin{align}
H_{\mathrm{int}}&=
V
\sum_{\langle\mathbf r,\mathbf r'\rangle,\sigma}
n_{\mathbf r\sigma}n_{\mathbf r'\sigma} + W
\sum_{\langle\mathbf r,\mathbf r'\rangle}
\left(
n_{\mathbf r\uparrow}n_{\mathbf r'\downarrow}
+
n_{\mathbf r\downarrow}n_{\mathbf r'\uparrow}
\right)
\label{eq:Hint}
\end{align}
where $n_{\mathbf r\sigma}=c^\dagger_{\mathbf r\sigma}c_{\mathbf r\sigma}$ and $V$ and $W$ are the repulsions between particles of equal and opposite spin, respectively. The interaction contains neither spin flips nor pair hoppings, and therefore conserves the number of particles of each spin, $N_\sigma$, or conserves $N_e$ and $S_z=(N_\uparrow-N_\downarrow)/2$. 


Let $\Phi_{\sigma,-}(\mathbf{k})$ be the lower band eigenvector of
Eq.~\eqref{eq:H_MF}, with eigenvalue $\lambda_{\mathbf{k}\sigma,-}$, and
define
\begin{align}
d^\dagger_{\mathbf{k}\sigma}
&\equiv \psi^\dagger_{\mathbf{k}\sigma}\Phi_{\sigma,-}(\mathbf{k})
\label{eq:projected_band_operator}
\end{align}
On a finite-sized torus, the band-projected many-body Hamiltonian becomes:
\begin{align}
\bar H=
\sum_{\mathbf{k},\sigma}
\lambda_{\mathbf{k}\sigma,-}
d^\dagger_{\mathbf{k}\sigma}d_{\mathbf{k}\sigma}
&+
\frac{1}{2N_s}
\sum_{\sigma,\sigma'}
\sum_{\{\mathbf{k}_i\}}
\mathcal M_{\sigma\sigma'}
(\mathbf{k}_1,\mathbf{k}_2;\mathbf{k}_3,\mathbf{k}_4)
\nonumber\\
&\quad\qquad\times
d^\dagger_{\mathbf{k}_1\sigma}
d^\dagger_{\mathbf{k}_2\sigma'}
d_{\mathbf{k}_4\sigma'}
d_{\mathbf{k}_3\sigma}
\label{eq:projected_ham}
\end{align}
where the second summation is restricted to processes that conserve total crystal
momentum (modulo a reciprocal vector of the RBZ). The explicit form of the projected interaction $\mathcal M_{\sigma\sigma'}$, together with details of the Fock space construction and exact diagonalization on a finite torus, is
given in Appendix~\ref{sec:projected_ED}.


In what follows, we first study the $W=0$ limit, where the Hamiltonian factorizes into two
independent spin sectors. This provides a basis for determining
how opposite-spin correlations reshape the many-body spectrum when $W$ is turned on. On its own, an apparent ground state many-body degeneracy at a single boundary condition at periodic boundaries is not sufficient to establish the existence of a fractionalized phase, so we also examine the manifold's response to spin-independent and spin-dependent flux insertion \cite{Sheng2006} and compute momentum occupation and charge- and spin-density correlations to check for possible competing orders.

\subsection{$\nu=\frac{1}{3}$}
\label{sec:nu_1_3}
 For the present calculations, we set $t=1$ eV as the energy scale and use the illustrative values $t'/t=0.11$, $t''/t=0.26$, $W_0/t=3.2$, and $\Delta_0/t=-1.65$. The corresponding band structure is shown in Fig.~\ref{fig:band}. Its valence bandwidth, $w\simeq0.24t$, is well below the direct band gap of approximately $3.3t$. We stress that this is a representative dispersive parameter set, rather than a uniquely optimized point or a fit to a particular material, and the qualitative many-body results persist under small parameter variations. Recall that the Bloch eigenvectors are solely determined by $t$, $W_0$, and $\Delta_0$, whereas $t'$ and $t''$ affect only the dispersion. Therefore, longer range hoppings or harmonics that are even under $\mathbf{k}\rightarrow\mathbf{k}+\mathbf Q$ may be considered to achieve even smaller bandwidths.

\begin{figure*}[t]
  \centering
  \includegraphics[width=\textwidth]{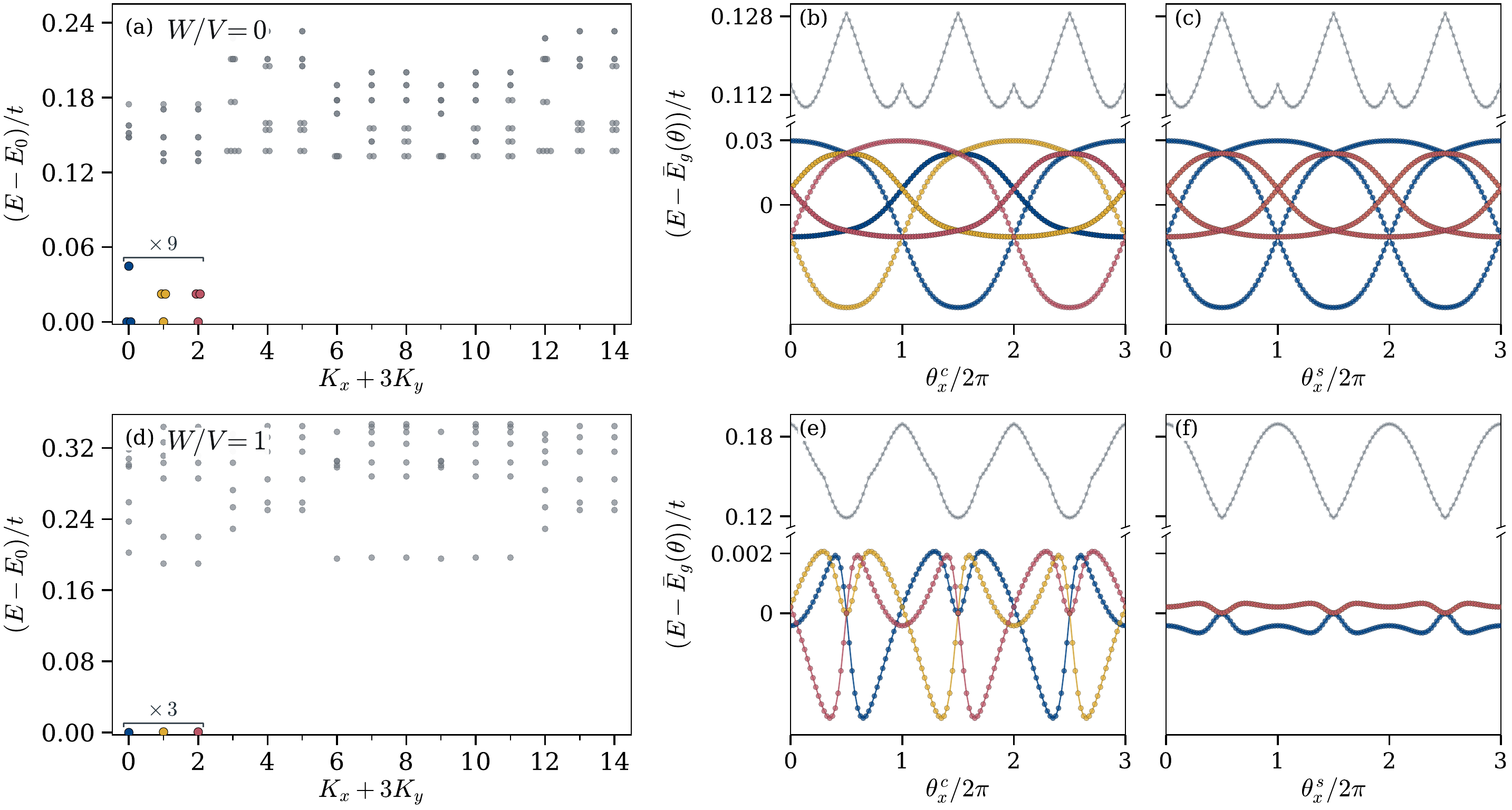}
  \caption{\textbf{Many-body spectra and flux response at $\nu=1/3$.} Exact-diagonalization results for $N_e=10$ particles on a $3\times5$ torus in the $S_z=0$ sector, with the same-spin repulsive interaction set to $V=1.5t$. Panels (a)--(c) show the decoupled limit $W=0$, while panels (d)--(f) show $W/V=1$. Panels (a) and (d) give the spectra at periodic boundary conditions as a function of a linearized momentum index $K=K_x+N_xK_y$, with energies measured from the lowest state. The nine-state product manifold at $W=0$ is replaced at $W/V=1$ by an isolated triplet in the momentum sectors $(0,0)$, $(1,0)$, and $(2,0)$. Panels (b) and (e) show spin-independent flux threading, while panels (c) and (f) show the spin-dependent one. In these panels, energies are reported relative to the instantaneous average of the ground state manifold, $\overline E_g(\theta)=g^{-1}\sum_{a=1}^{g}E_a(\theta)$, with $g=9$ for $W=0$ and $g=3$ for $W/V=1$.}
\label{fig:nu_1_3_spectra}
\end{figure*}

\paragraph{The spin-decoupled limit.} We first set the opposite-spin interaction $W$ to zero while keeping the same-spin repulsion at $V=1.5t$. We consider a $3\times5$ torus with $N_\uparrow=N_\downarrow=5$, so each of the lower bands is 1/3 filled. In this limit, the Hamiltonian factorizes as $H=H_\uparrow+H_\downarrow$, and a complete many-body eigenbasis can be written in a product form $|\Psi_{ab}\rangle=|\psi_{a,\uparrow}\rangle\otimes|\psi_{b,\downarrow}\rangle$, with $E_{ab}=E_{a,\uparrow}+E_{b,\downarrow}$. The many-body spectrum for each spin component contains three low-lying states, one in each of the momentum sectors $(K_x,K_y)=(0,0),(1,0),(2,0)$. This is the torus counting consistent with a $\nu=1/3$ Laughlin-like state of a fractional Chern insulator \cite{Regnault2011}. Because total momentum is obtained by adding the two spin-sector momenta modulo $(N_x,N_y)$, the product wavefunction has nine low-lying states, with three states in each of the same three momentum sectors. Such a ninefold product manifold is the expected decoupled limit of a spin-conserving FQSH state, which in the continuum is described by two, time-reversed Laughlin states, with holomorphic and anti-holomorphic dependence on the complex coordinates \cite{Bernevig2006}. Fig.~\ref{fig:nu_1_3_spectra} shows the many-body spectrum for this case. The low-energy quasidegenerate manifold has a spread of $\delta_9=E_9-E_1 \simeq0.045t$, and a gap of $\Delta_9=E_{10}-E_9\simeq0.084t$ sits above it.

We next test whether this quasidegenerate manifold remains isolated when threading flux, which we implement numerically through twisted boundary conditions in the doubled unit cell coordinates. For a twist in the $x$ direction, $c_{\mathbf r+N_x\hat{\mathbf x},\sigma}=e^{i\theta_{x,\sigma}}c_{\mathbf r,\sigma}$, we write the spin-dependent twist angles as $\theta_{x,\uparrow}=\theta_x^c+\theta_x^s$ and $\theta_{x,\downarrow}=\theta_x^c-\theta_x^s$. A spin-independent flux, $\theta_x^s=0$, applies the same phase to the two spins, whereas a spin-dependent flux, $\theta_x^c=0$, applies opposite phases to the spin $\uparrow$ and $\downarrow$ electrons. One flux quantum corresponds to increasing the relevant twist angle by $2\pi$. To better visualize the internal spectral flow, Figs.~\ref{fig:nu_1_3_spectra}(b) and \ref{fig:nu_1_3_spectra}(c) plot the energies relative to the instantaneous average of the 9 state manifold, $E_a(\theta)-\overline E_9(\theta)$, where $\overline E_9(\theta)=\sum_{a=1}^{9}E_a(\theta)/9$. Under both twists, the lowest nine states remain distributed as three states in each of $(0,0),(1,0),(2,0)$ and stay separated from the higher spectrum throughout the cycle.


Although both spectra coincide as unordered energy sets at $W=0$, their momentum-sector mappings differ. Writing the inserted flux in units of the flux quantum as $\Phi=\theta_x/2\pi$ and labeling the three levels $r$ in sector $K_x$ by $E_{K_x,r}$, the spin-independent twisted spectra obey $\{E^c_{K_x,r}(\Phi+1)\}_r=\{E^c_{K_x+1,r}(\Phi)\}_r$, with momentum understood mod $N_x$. In our convention, this corresponds to $K_x\rightarrow K_x-1$, so the spectrum returns to itself after three charge flux quanta. For spin-dependent flux threading, the relation is instead $\{E^s_{K_x,r}(\Phi+1)\}_r=\{E^s_{K_x,r}(\Phi)\}_r$. The net momentum shift vanishes because $N_\uparrow-N_\downarrow=0$. The crossings in Fig.~\ref{fig:nu_1_3_spectra}(c) reflect changes in the energy ordering within this ground state manifold through the flux cycle.

\paragraph{The $W\neq0$ case.} We now turn on the opposite-spin repulsion $W$ while keeping $V=1.5t$. The Hamiltonian no longer separates into independent spin sectors, although $N_\uparrow$, $N_\downarrow$, and $S_z$ remain conserved. In momentum space, the new interaction transfers equal and opposite crystal momentum between the two spin species and mixes the states within the ninefold product manifold as the interspin coupling is slowly turned on. At $W/V=0.10$, the nine states split into a lower triplet and an upper sextet, but all nine remain isolated from the higher spectrum. The 9 state gap decreases further with $W$ and remains finite until $W/V=0.30$, where the boundary between the ninth and tenth states closes. Beyond $W/V \geq 0.40$, states from other momentum sectors enter the lowest nine levels, so the original manifold is no longer isolated, while the lowest triplet remains separated from the rest of the spectrum.

At $W=V$, the interaction reduces to the spin-independent form $V\sum_{\langle\mathbf r,\mathbf r'\rangle}
n_{\mathbf r}n_{\mathbf r'}$, where $n_{\mathbf r}=n_{\mathbf r\uparrow}+n_{\mathbf r\downarrow}$. Figure~\ref{fig:nu_1_3_spectra}(d) shows the spectrum at $W/V=1$, where $V=W=1.5t$. The three lowest states occur in the same momentum sectors as before, with the states in $(1,0)$ and $(2,0)$ being degenerate to numerical precision. Their spread is much smaller, $\delta_3=E_3-E_1\simeq6.3\times10^{-4}t$, while the gap above the triplet is substantially larger, $\Delta_3=E_4-E_3\simeq0.189t$. This shows that the triplet is very well isolated at periodic boundary conditions.


Panels~\ref{fig:nu_1_3_spectra}(e) and \ref{fig:nu_1_3_spectra}(f) show the response of this triplet to charge and spin flux, respectively. Under charge flux, the lowest three states continue to contain one state in each of the momentum sectors. The three sectors undergo the same cyclic, large gauge transformation found in the decoupled limit and return to their initial configuration after three charge flux quanta. Under spin flux, the triplet remains even more tightly grouped, and states in $(1,0)$ and $(2,0)$ remain degenerate throughout the path, while each momentum sector returns to itself after one spin flux quantum. The flux calculations therefore show that the triplet remains stable under both common and relative boundary twists, with sector cycling occurring only under charge flux. 

\begin{figure}[!t]
\centering
\includegraphics[width=0.78\columnwidth]{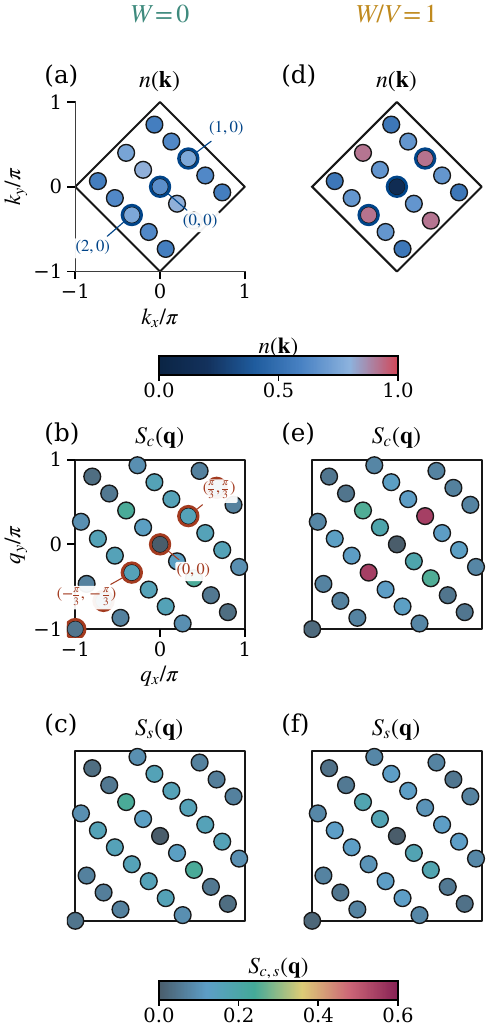}
\caption{\textbf{Momentum occupation and density correlations at $\nu=1/3$.} Equal-weight quasidegenerate manifold averages at $W=0$ [panels (a)--(c)] and $W/V=1$ [panels (d)--(f)]: momentum occupation $n(\mathbf{k})$ on the RBZ [(a),(d)], charge structure factor $S_c(\mathbf{q})$ [(b),(e)], and spin structure factor $S_s(\mathbf{q})$ [(c),(f)] of the 30 microscopic sites.}
\label{fig:nu_1_3_correlations}
\end{figure}

The persistence of the triplet under flux insertion establishes a robust ground state manifold, but does not distinguish a homogeneous FQSH liquid from a state with broken translation symmetry~\cite{Regnault2011}. To address this possibility, Fig.~\ref{fig:nu_1_3_correlations} compares the momentum occupation and static density correlations in the decoupled and coupled limits. We average the observables over the ground state manifold using
$\hat\varrho_g=g^{-1}\sum_{a=1}^{g}|\Psi_a\rangle\langle\Psi_a|$,
with $g=9$ at $W=0$ and $g=3$ at $W/V=1$. Panels (a) and (d) show the momentum occupation
\begin{align}
n(\mathbf{k})&=
\sum_\sigma
\operatorname{Tr}\!\left[
\hat\varrho_g
d^\dagger_{\mathbf{k}\sigma}d_{\mathbf{k}\sigma}
\right]
\end{align}
Panels (b), (c), (e), and (f) show the charge and spin structure factors for both cases,
\begin{align}
S_\alpha(\mathbf{q})
&=\frac{1}{N_s}\Big[
\langle\bar\rho_\alpha(-\mathbf{q})
\bar\rho_\alpha(\mathbf{q})\rangle_g-\langle\bar\rho_\alpha(-\mathbf{q})\rangle_g
\langle\bar\rho_\alpha(\mathbf{q})\rangle_g
\Big]
\label{eq:connected_structure_factor}
\end{align}
where $\alpha=c,s$, $\langle\mathcal O\rangle_g=\operatorname{Tr}(\hat\varrho_g\mathcal O)$, and $N_s=30$ is the number of microscopic lattice sites. Here $\bar\rho_c=\bar\rho_\uparrow+\bar\rho_\downarrow$, $\bar\rho_s=\bar\rho_\uparrow-\bar\rho_\downarrow$, and $\bar\rho_\sigma(\mathbf{q})$ is the physical microscopic density projected into the lower stDDW band. At $W=0$, the occupation is distributed nearly uniformly over the entire RBZ, while $S_c(\mathbf{q})=S_s(\mathbf{q})$ and the cross-spin correlations vanish, as expected of two decoupled $\nu=1/3$ liquids, i.e., each spin sector develops its own correlations, but their density fluctuations remain uncorrelated.

The opposite-spin interaction changes this behavior qualitatively. At $W=V$, the nearest-neighbor interaction is spin independent and couples directly to the total charge density. Correspondingly, the charge structure factor in Fig.~\ref{fig:nu_1_3_correlations}(e) develops a sharp peak at $\mathbf{q}_\ast=\pm(\pi/3,\pi/3)$, with $S_c(\mathbf{q}_\ast)=0.546$, whereas no analogous peak appears in the spin response, for which $S_s(\mathbf{q}_\ast)=0.098$. The cross-spin structure factor $S_{\uparrow\downarrow}$ is defined using the same normalization as Eq.~\eqref{eq:connected_structure_factor}. Since
$S_c(\mathbf{q})-S_s(\mathbf{q})=4\,\mathrm{Re}\,S_{\uparrow\downarrow}(\mathbf{q})$, the difference at $\mathbf{q}_\ast$ is a positive interspin correlation.
At the same time, the momentum occupation becomes strongly nonuniform, with its relative variation increasing from $0.127$ at $W=0$ to $0.305$ at $W/V=1$. Because the single-particle Hamiltonian is unchanged, both effects are induced by the opposite-spin interaction. We do not seek to establish spontaneous translation-symmetry breaking from a single finite-size cluster, but the combination of a sharply selected charge wave vector, the absence of a corresponding spin peak, and the increasingly nonuniform momentum occupation favors a charge-ordered state (period three CDW) interpretation of the coupled triplet over a homogeneous ``paired'' FQSH liquid.

\paragraph{Skyrmion charge.}
Before proceeding to the discussion of numerical results at different fillings, we briefly pause to discuss an intriguing possibility associated with such a broken symmetry parent state. Recall that the orientation $\hat{\mathbf n}$ of the triplet order parameter can vary slowly. Suppose that the decoupled $\nu=1/m$ FQSH phase survives in an $SU(2)$-symmetric continuation. Let $A^c$ be an external probe field coupled to the total particle number current. Another field, $A^s$, couples to $S_z$ and, for a slowly varying texture, is the spin connection generated when the local spin axis is rotated to follow $\hat{\mathbf n}$. The two spin sectors couple as $\eta_\sigma A^s/2$ and have opposite Chern numbers, $C_\sigma=\eta_\sigma C$, where $|C|=1$. The real time, long-wavelength response is:

\begin{align} S &=- \sum_\sigma \frac{C_\sigma}{4\pi m} \int \left(A^c+\frac{\eta_\sigma}{2}A^s\right) \wedge d\left(A^c+\frac{\eta_\sigma}{2}A^s\right) \nonumber\\ &=- \frac{C}{2\pi m}\int A^c\wedge dA^s \end{align} 
where $\wedge$ denotes the wedge product. In component notation, $\int A^c\wedge dA^s = \int d^3x \epsilon^{\mu\nu\lambda} A_\mu^c\partial_\nu A_\lambda^s$.


The pure charge and pure spin terms cancel between the two time-reversed sectors, leaving only the mixed response. For a skyrmion of Pontryagin index $\chi$ whose size is large compared with the electronic correlation length, the spin gauge flux is $\int dA^s=4\pi\chi$~\cite{GroverSenthil2008,HsuChakravarty2013}. Its spatial winding acts as opposite effective fluxes of $2\pi\chi$ and $-2\pi\chi$, for the two spin sectors. Since their Chern numbers are also opposite, the two sectors contribute the same particle number and the contributions add. The mixed response gives the particle number and electric charge bound to an adiabatically formed texture as \begin{align} \Delta N_{\mathrm{sk}} &= \frac{C}{2\pi m}\int dA^s = \frac{2C\chi}{m}, \qquad Q_{\mathrm{sk}} = -\frac{2eC\chi}{m} \end{align} 


The two spin sectors contribute equally, $\Delta N_\uparrow=\Delta N_\downarrow=C\chi/m$, so this response assigns no net $S_z$ to the texture. For $|C|=|\chi|=1$, a unit skyrmion has charge of magnitude $2e/m$, or $2e/3$ at $m=3$. In such a scenario, the fractional charge is attached to a topological texture of the order parameter or collective coordinate that generates the Chern bands. 


In our numerics, the nine-state manifold occurs at the spin-anisotropic point $W=0$, whereas the point $W=V$, at which the interaction is spin independent, seems to exhibit strong period-three charge correlations. The present calculations therefore do not establish that the FQSH phase survives when $\hat{\mathbf n}$ is allowed to fluctuate. An interesting possibility is to study whether extended interactions that preserve spin-rotation symmetry can stabilize a FQSH phase and determine the skyrmion energy and stability once the dynamics of $\hat{\mathbf n}$ are restored.


\subsection{$\nu=\frac{1}{5}$}
\begin{figure}
        \centering
        \includegraphics[width=0.8\columnwidth]{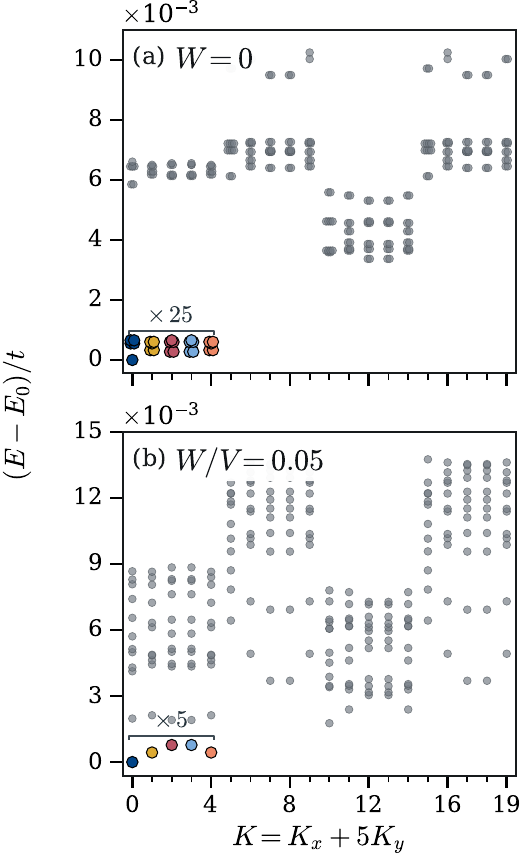}
        \caption{\textbf{Many-body spectra at $\nu=1/5$.} Exact diagonalization results for a $5\times4$ torus ($N_\uparrow=N_\downarrow=4$). Panels (a) and (b) show the spectra at $W=0$ and $W/V=0.05$, respectively. Colored circles denote the lowest 25 states at $W=0$ and the lowest 5 states at $W/V=0.05$.}
\label{fig:nu_1_5_spectra}
\end{figure}

We next ask whether a FQSH phase can be stabilized at the more dilute filling $\nu=1/5$. Unlike the $\nu=1/3$ case, we observe that this filling requires much more stringent conditions: the spectra shown in Fig.~\ref{fig:nu_1_5_spectra} are obtained by setting $t'$ and $t''$ to zero and working in the flat-band limit, i.e., a projection that does not change the stDDW wave functions but removes their single-particle energy dispersion. The question, however, remains whether the stDDW Bloch states can support a fractional state. The answer is yes, although only on a much smaller energy scale. At $W=0$, the two spin sectors are decoupled, each containing five quasidegenerate ground states, or $5^2$ product states in the full system. On a $5\times4$ torus, these states have a small spread and are separated from higher states by a gap of $\Delta_{25}$ = 2.7 meV. Indeed, the stDDW Bloch states do support the counting expected of two decoupled $\nu=1/5$ Laughlin states, albeit much more fragilely.

Because of this, the state is also very susceptible to interspin coupling. Indeed, at $W/V=0.05$ (corresponding to $W \gg \Delta_{25}$) the many-body spectrum is reorganized into five states with a spread $\delta_5=7.8\times10^{-4}t$ and gap of $\approx$ 1 meV above them. By $W/V=1$, this manifold disappears completely and the system has a unique ground state. These results should be contrasted with the $\nu=1/3$ case above, where a ground state many-body degeneracy is not vitiated by band dispersion. Supporting calculations on a $5\times3$ torus lead to similar conclusions.

\section{Other fillings}
\subsection{$\nu=\frac{1}{7}$}
At even more dilute fillings such as $\nu=1/7$, none of the computationally tractable systems show a ground state pattern consistent with a FQSH phase, regardless of how dispersive the band is. For the same parameters we used for $\nu=1/3$, a $3\times7$ torus (corresponding to $N_\uparrow=N_\downarrow=3$), for example, has a fourfold degenerate ground state at $W=0$ and a unique ground state at $W/V=1$. The momentum occupation for this system is strongly nonuniform, while its charge structure factor shows modest  enhancements at wave vectors along the BZ diagonals, $q_x=-q_y$ (See Appendix Fig.~\ref{fig:nu_1_7_spectra_correlations}). Though they identify the dominant short-range correlations, one needs to scale to larger system sizes to more carefully establish whether these features are Bragg peaks associated with a crystalline state    such as a CDW or a Wigner crystal.

\subsection{$\nu=\frac{1}{2}$}

Before concluding, we discuss what possible phases the system may realize if each of the lower bands is half-filled. Even-denominator fillings are characteristically different from the odd-denominator $\nu=1/m$ Laughlin states whose zero-field lattice analogues we have discussed so far. Most famously, the half-filled, spin-polarized lowest Landau level (LLL) has no quantized Hall plateau \cite{Jiang1989} and is described in the HLR theory~\cite{HalperinLeeRead1993} as a compressible, composite Fermi liquid~\cite{Jain1989,Lopez1991,Son2015}. Here, we use the same interactions and single-particle parameters as in the $\nu=1/3$ case. But before considering the many-body problem, it is instructive to begin with the noninteracting case as a reference and then see how repulsive interactions renormalize it.

\begin{figure}[t]
\centering
\includegraphics[width=0.96\columnwidth]{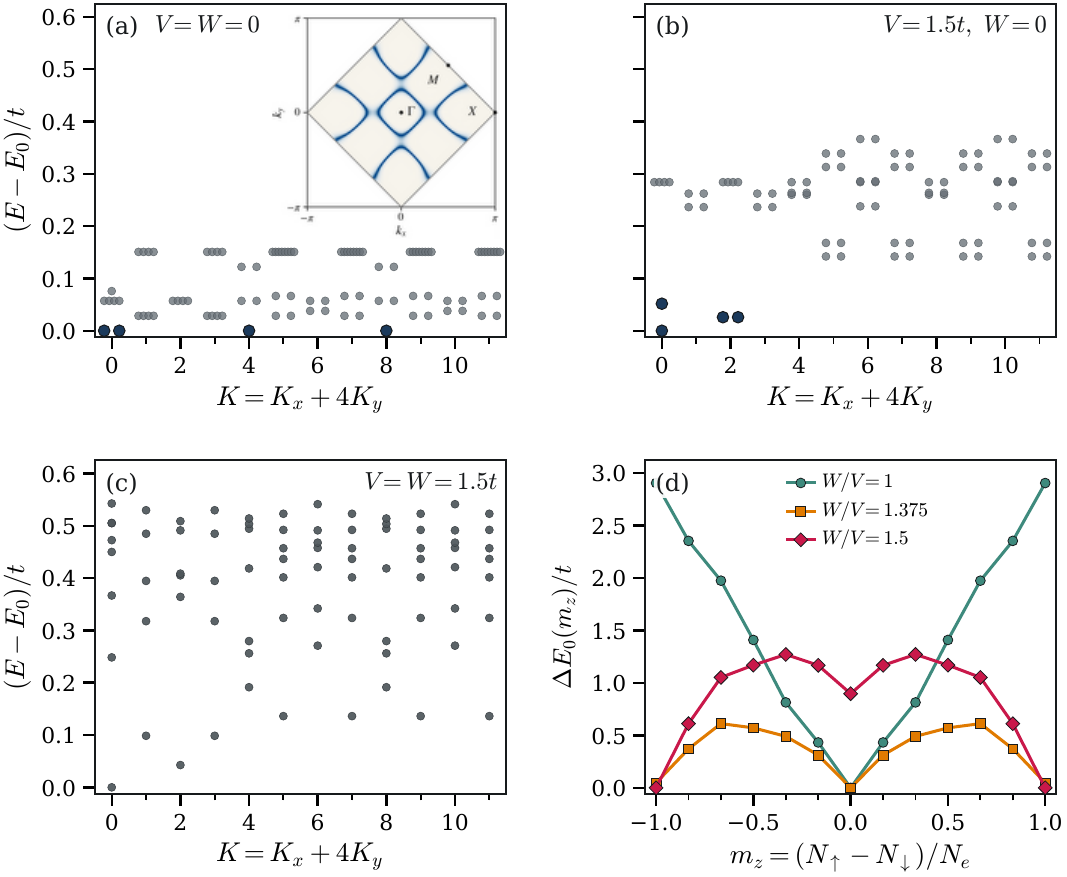}
\caption{\textbf{stDDW spectra at $\nu=1/2$.} Exact-diagonalization results for $N_\sigma=6$ particles on a $4\times3$ torus. Panels (a)--(c) show the $S^z=0$ spectra at $V=W=0$, $V=1.5t$ and $W=0$, and $V=W=1.5t$, respectively; the inset in panel (a) shows the hole-like Fermi surface, and the spectrum shows excitations of the two independent noninteracting Fermi seas. Dark blue points represent four exactly degenerate ground states in panel (a) and the four quasidegenerate states in (b), whereas the lowest levels are not isolated from their excitations in panel (c). Panel (d) shows the ground state energy in each spin-polarization sector, measured relative to the lowest sector energy at the same value of $W/V$, for $W/V=1$, $1.375$, and $1.5$. The minimum moves from $m_z=0$ at $W/V=1$ to the two $m_z=\pm1$ endpoints at $W/V=1.5$, which form a time-reversal pair. At $W/V=1.375$, the unpolarized state and polarized endpoints differ only by $\approx 0.05t$.}
\label{fig:nu_1_2_spinful_spectra}
\end{figure}

At half filling of the lower bands and $V=W=0$, the stDDW model is a metal, invariant under time reversal, with vanishing charge Hall conductivity and finite but unquantized intrinsic spin Hall conductivity. To make contact with the finite-size exact diagonalization numerics, we plot the particle-hole excitations of its Fermi seas on a finite $4\times 3$ torus with $N_\sigma=6$, where the allowed momenta are discrete and six of the twelve orbitals are occupied for each spin. The resulting spectrum in Fig.~\ref{fig:nu_1_2_spinful_spectra}(a) shows an exact fourfold ground state degeneracy, corresponding to two degenerate ground states per spin. However, this multiplicity is an artifact of the discrete momentum grid and changes with the system size and geometry (e.g., 36-fold on $4\times4$ and 4-fold on $5\times4$).


When same-spin repulsive interactions ($V=1.5t$) are turned on, the interacting spectrum shows a fourfold quasidegeneracy, Fig.~\ref{fig:nu_1_2_spinful_spectra}(b). As in the noninteracting case, however, the multiplet is strongly sensitive to the cluster geometry. Furthermore, the spectrum has a charge gap that remains small on larger tori, although it does not decrease monotonically, and the quasidegenerate manifold merges with higher states under flux insertion. The absence of a persistent topological ground state with a gap above it therefore disfavors the possibility that the system realizes an incompressible phase with the chosen interactions, such as a fermionic Moore-Read state~\cite{Rezayi1994}. Our calculations for several different cluster sizes also show no sign of interaction-driven crystallization. As the data in Fig.~\ref{fig:nu_1_2_correlations_all_geoms} show, the largest peak in the connected structure factor remains of order one rather than growing with the system size, while $n(\mathbf{k})$ becomes smoother and averages to about one half throughout the Brillouin zone.


What then is the nature of this homogeneous liquid? Are its low-energy excitations electrons or composite fermions? Two hints suggest it is the latter. The first comes from the band-projected electron spectral function. Acting with $d_{\mathbf{k}}^\dagger$ or $d_{\mathbf{k}}$ on a Slater determinant produces one exact eigenstate whenever the corresponding addition or removal process is allowed. The corresponding spectral function for the $V=0$ case should therefore contain one $\delta$-function peak with unit quasiparticle weight. This changes dramatically when interactions are turned on. In the $V=1.5t$ case, adding or removing an electron produces a superposition of many interacting eigenstates, and the spectral weight is fragmented among many peaks. To reduce the dependence on the discrete momentum grid, we average the spectral function over generic twist angles and plot the spectra in Fig.~\ref{fig:nu_1_2_spectral_fun}, but we analyze the pole weights separately at each momentum and twist angle. For example, on the 24 orbital tori ($3\times 8$ or $6 \times 4$), the largest pole is estimated to carry only 0.13-0.17 of the total addition or removal weight at that momentum. The corresponding participation ratio shows that this weight is effectively shared among 17-21 excited states [Fig.~\ref{fig:nu_1_2_spectral_weights}]. Although the finite-size exact diagonalization data are insufficient to establish that the electron quasiparticle weight vanishes in the thermodynamic limit, they show a significant loss of electron coherence as the system size increases. This behavior is more consistent with a composite Fermi liquid than with a weakly renormalized electron metal \cite{Pichler2025}. The second hint comes from comparison with the half-filled, spin-polarized LLL. Same-spin repulsive interactions renormalize the lowest energies across momentum sectors toward a pattern that is reminiscent of a compact composite fermion Fermi sea, whereas the noninteracting spectrum lacks such splittings. However, the detailed one to one level ordering and ground state momenta are not matched on every torus (see Fig.~\ref{fig:nu_1_2_lll_comparison} for comparison across different cluster geometries).


Lastly, when $W$ is turned on, it couples the two spin sectors, so the product-state interpretation is no longer applicable. At $W/V=1$, the ground state remains unpolarized and is not well-separated from higher excitations, Fig.~\ref{fig:nu_1_2_spinful_spectra}(c). 
What the precise identity of this correlated phase is remains not straightforward to answer and requires a much more sophisticated treatment. The situation, however, becomes unambiguous at larger $W$. In Fig.~\ref{fig:nu_1_2_spinful_spectra}(d), we calculate the ground state energy for each allowed spin $S_z$ projection and plot it against spin polarization, $m_z=2S_z/N_e$, where $m_z=0$ is unpolarized and $m_z=\pm1$ is fully polarized. Clearly, between $W/V\approx 1.4$ and 1.5, the system undergoes a first order transition to fully polarized sectors, without any partially polarized sector becoming the ground state. In this regime, one of the lower Chern bands is completely filled while its time-reversed partner is empty and the model realizes a time reversed pair of quantum anomalous Hall ferromagnets with Chern numbers $C=\pm1$.

\section{Conclusion}
\label{sec:conclusion}

We have studied whether the FQSH effect can emerge from a symmetry broken parent state in which a mixed singlet--triplet $d$-density wave produces a pair of Chern bands that are related by time-reversal symmetry. At integer filling, the parent state realizes an IQSH insulator, whereas at fractional filling, its response depends sensitively on the balance between the electrons' kinetic energy and their repulsive interactions. At $\nu=1/3$, the decoupled limit gives the expected ninefold-degenerate ground state for the product state. Opposite-spin interactions reduce this manifold into a triplet, but the accompanying enhancement of charge correlations indicates that the strongly coupled state is likely not a homogeneous FQSH liquid. At $\nu=1/5$, the expected many-body ground state degeneracy appears only in the flat-band limit, showing that the stDDW Bloch states can support this more fragile fraction, while the physical dispersion spoils it. 


No corresponding fractional manifold appears at $\nu=1/7$ on the accessible clusters. Across these odd-denominator fillings, the resulting hierarchy resembles that of the charge quantum Hall problem, where low-denominator fractional liquids are the most robust, while lower fillings and residual dispersion increasingly favor competing ordered states. The fact that the $\nu=1/7$ spectra show no evidence of a FQSH phase is not entirely surprising, considering that the charge quantum Hall problem suggests competition between liquid and crystalline states at such a dilute filling. In our model, the modest enhancement of charge correlations and strongly nonuniform momentum occupation are consistent with such a tendency, but are insufficient to establish a CDW or Wigner crystal. Moreover, because the stDDW parent already breaks translation symmetry in a hidden bond channel, the competing state may involve additional charge, spin, or bond order rather than a Wigner crystal. Resolving this question requires larger systems and finite-size scaling of the relevant correlations. 


At half-filling, the fragmentation of spectral weight and the evolution of momentum sector energies across different system sizes at $W=0$ suggest a compressible state, likely a time-reversed pair of composite Fermi liquids. The coupled state at $W=V$ may also be compressible but still requires a careful identification. Upon further increasing the opposite-spin repulsion, the system undergoes a direct level crossing into the fully polarized quantum anomalous Hall ferromagnets.


An interesting feature of the spin Hall response is that its natural quantum contains no explicit $\hbar$. For either spin sector, the electrical Hall conductance is quantized in units of $e^2/h$. An electron carries charge of magnitude $e$ and spin projection $\hbar/2$, so within a given sector, its spin current is obtained by multiplying its charge current by $(\hbar/2)/e$. A single sector therefore contributes a spin Hall conductance of magnitude $(\hbar/2e)(e^2/h)=e/4\pi$. For the time-reversed pair, both the Hall chirality and the spin polarization are reversed. The charge Hall currents consequently cancel, whereas the spin Hall currents add, giving $|\sigma_{xy}^{\mathrm{spin}}|=\nu (e/2\pi)$ \cite{Bernevig2006}. The cancellation of $\hbar$ reflects the mixed nature of the response in which the electric field couples to charge, whereas the transverse current carries spin. Its exact quantization requires a conserved spin component. In the spin-rotation invariant stDDW Hamiltonian, the triplet condensate selects an axis but preserves rotations about it. Choosing this direction as the $z$ axis identifies the unbroken $U(1)$ subgroup with conservation of $S_z$.


Finally, the effect of disorder was not included in the present calculations. In the quantum Hall effect, disorder localizes excitations and allows quantized plateaus to persist over a finite range of filling, whereas sufficiently strong disorder destroys the mobility gap; see, however, an alternate view expressed in Ref.~\cite{Kim2021}. An additional distinction arises here because exact quantization of the spin Hall response requires conservation of $S_z$: scalar and spin-dependent disorder therefore need not have the same effect. Determining the disorder dependence of this hidden density-wave order and its fractional phases is an interesting direction for future work.
                       
\section{Acknowledgments}
We thank Srinivas Raghu and Hong-Wen Jiang for discussion.
This work used computational and storage services associated with the Hoffman2 Cluster which is operated by the UCLA Office of Advanced Research Computing’s Research Technology Group.

\appendix
\counterwithin{figure}{section}
\section{Chern numbers and Hall conductances}
\label{app:hall}

We calculate the Hall response when the lower band of each spin sector is
filled. The electron charge is taken to be $-e$ and the RBZ is oriented by $dk_x\wedge dk_y$. We  define
$\mathcal A_{\sigma,i}
\equiv i\langle u_{\sigma,-}(\mathbf{k})|\partial_{k_i}u_{\sigma,-}(\mathbf{k})\rangle$.
The spectral representation of the Kubo formula expresses the Berry
curvature of the lower band, $\Omega_{\sigma,-}
=\partial_{k_x}\mathcal A_{\sigma,y}
-\partial_{k_y}\mathcal A_{\sigma,x}$, as follows, where common $\mathbf{k}$ and
$\sigma$ labels are suppressed inside the matrix elements and eigenvalues:
\begin{align}
\Omega_{\sigma,-}(\mathbf{k})
&=-2\,\mathrm{Im}\,
\frac{\langle -|\partial_{k_x}h_\sigma|+\rangle
\langle +|\partial_{k_y}h_\sigma|-\rangle}
{\left(\lambda_{-}-\lambda_{+}\right)^2}
\nonumber\\
&=\frac{\vec h_\sigma\cdot
\left(\partial_{k_x}\vec h_\sigma\times
\partial_{k_y}\vec h_\sigma\right)}
{2E_{\mathbf{k}}^{3}}
\label{eq:berry_two_band}
\end{align}
Here $\vec h_\sigma=(\Delta_{\mathbf{k}},-\eta_\sigma W_{\mathbf{k}},
\varepsilon_{1\mathbf{k}})$. The scalar term
$(\varepsilon_{2\mathbf{k}}-\mu)\tau_0$ has no interband matrix elements and
does not change the eigenvectors. It can nonetheless determine when the indirect gap closes. This is
why $t'$ and $t''$ do not appear below.


Substituting the form factors from Eq.~\eqref{eq:form_factors} into
Eq.~\eqref{eq:berry_two_band} gives
\begin{align}
\Omega_{\sigma,-}(\mathbf{k})
&=\eta_\sigma\frac{tW_0\Delta_0}{E_{\mathbf{k}}^{3}}
\left(\sin^2k_y+\sin^2k_x\cos^2k_y\right)
\end{align}
The Chern number is
$C_\sigma=(2\pi)^{-1}\int_{\mathrm{RBZ}}d^2k\,
\Omega_{\sigma,-}(\mathbf{k})$. The two spin sectors have opposite Berry
curvature, and if either component of the mixed
density wave is zero, the direct gap closes. Both the triplet and singlet orders are therefore 
needed for the topological state.


The Chern number carried by each band can be most transparently seen in the
small-$|\Delta_0|$ limit. When $\Delta_0=0$, the folded spectrum has two Dirac points in the RBZ,
$\mathbf K_1=(\pi/2,\pi/2)$ and
$\mathbf K_2=(\pi/2,-\pi/2)$. Writing
$\mathbf{k}=\mathbf K_i+\mathbf{q}$ and linearizing in $\mathbf{q}$ (to  leading order):
\begin{align}
 h_{\sigma}^{(1)}&=\Delta_0\tau_1
+\eta_\sigma\frac{W_0}{2}(q_x-q_y)\tau_2
+2t(q_x+q_y)\tau_3
\nonumber\\
h_{\sigma}^{(2)}&=-\Delta_0\tau_1
+\eta_\sigma\frac{W_0}{2}(q_x+q_y)\tau_2
+2t(q_x-q_y)\tau_3
\end{align}
The $d_{xy}$ order provides the ``masses''
$m_1=\Delta_0$ and $m_2=-\Delta_0$. The corresponding 
chiralities of the Dirac points are determined by the signs of the Jacobians,
$J_i=\det[\partial(d_2,d_3)/\partial(q_x,q_y)]$, with
$J_1=2\eta_\sigma tW_0$ and
$J_2=-2\eta_\sigma tW_0$


For a single massive Dirac cone
$h_{\text{Dirac}}=m\tau_1+d_2(\mathbf{q})\tau_2+d_3(\mathbf{q})\tau_3$, with both
$d_2$ and $d_3$ linear in $\mathbf{q}$, the occupied band contributes
\begin{align}
C_{\text{Dirac}}
&=\frac{1}{2\pi}\int \mathrm d^2q\,
\frac{mJ}{2(m^2+d_2^2+d_3^2)^{3/2}}
\nonumber\\
&=\frac{\operatorname{sgn}J}{4\pi}
\int_0^{2\pi} d\varphi
\int_0^\infty  dQ\,
\frac{mQ}{(m^2+Q^2)^{3/2}}
\nonumber\\
&=\frac{1}{2}\operatorname{sgn}(mJ)
\end{align}
where in the second equality we used $d^2q=dd_2dd_3/|J|$ and changed to polar coordinates.
Between the two nodes, both the mass and the chirality reverse.
Consequently $m_1J_1$ and $m_2J_2$ have the same sign, so the two
half-integer contributions add:
\begin{align}
C_\sigma
&=\eta_\sigma\operatorname{sgn}(tW_0\Delta_0),
\qquad
C_\uparrow+C_\downarrow=0
\end{align}
Because the direct gap remains open for $tW_0\Delta_0\neq 0$, this result is valid away from the small-$|\Delta_0|$
limit. 


At zero temperature, with the chemical potential lying in the indirect bulk gap, each
filled spin block has the electrical Hall response $j_{x, \sigma}^{\mathrm{charge}}=\sigma^{(\sigma)}_{xy}E_y$, with
$\sigma^{(\sigma)}_{xy}=-(e^2/h)C_\sigma$. We let $j_{x,\sigma}$ denote the particle current in that spin sector, such that $j^{\mathrm{charge}}_{x,\sigma}=-e j_{x,\sigma}$ and define
$j_x^{\mathrm{spin}}
=(\hbar/{2})(j_{x,\uparrow}-j_{x,\downarrow})
=\sigma_{xy}^{\mathrm{spin}}E_y$. This gives the Hall conductances:
\begin{align}
\sigma^{\mathrm{charge}}_{xy}
&=\sigma^{(\uparrow)}_{xy}+\sigma^{(\downarrow)}_{xy}=0,\nonumber\\
\sigma^{\mathrm{spin}}_{xy}
&=-\frac{\hbar}{2e}
\left(\sigma^{(\uparrow)}_{xy}-\sigma^{(\downarrow)}_{xy}\right)
=\frac{e}{2\pi}\operatorname{sgn}(tW_0\Delta_0)
\end{align}
Note that the sign of the spin Hall conductance is determined relative to the orientation of $dk_x\wedge dk_y$, and the conventional choices of $j_x=\sigma_{xy}E_y$ and the spin quantization axis.

\section{Projected many-body Hamiltonian and its exact diagonalization}
\label{sec:projected_ED}

Here we delve into the details of the numerical diagonalization of the many-body Hamiltonian. We work on a rectangular torus of dimensions $N_x\times N_y$ and first impose periodic boundary conditions (PBCs). The allowed momenta are
\begin{align}
\mathbf{k}_{\mathbf n}
&=\frac{n_x}{N_x}\mathbf b_x
+\frac{n_y}{N_y}\mathbf b_y,
\nonumber\\
n_\alpha&=0,\ldots,N_\alpha-1,
\qquad \alpha=x,y
\end{align}
where $\mathbf b_x$ and $\mathbf b_y$ are the reciprocal lattice vectors. Each of the spin-up and spin-down bands contains $N=N_xN_y$
single-particle states, so the Kramers' pair contains $2N$ states.


As we mentioned above, we take $\Phi_{\sigma}(\mathbf{k})$ to denote the lower band eigenvector of Eq.~\eqref{eq:H_MF}, $\Phi_{\sigma}(\mathbf{k})=\left(u_{\sigma,-}(\mathbf{k}),v_{\sigma,-}(\mathbf{k})\right)^T$, with eigenvalue $\lambda_{\mathbf{k}\sigma,-}$. In the folded basis, $\psi^\dagger_{\mathbf{k}\sigma}
=
\big(
c^\dagger_{\mathbf{k}\sigma},
c^\dagger_{\mathbf{k}+\mathbf Q,\sigma}
\big),$ the band-projected creation operator is defined in Eq.~\eqref{eq:projected_band_operator}.
Projecting into the time-reversal pair therefore amounts to replacing $c_{\mathbf{k}\sigma}$ and $c_{\mathbf{k}+\mathbf Q,\sigma}$ with $u_{\sigma,-}(\mathbf{k})d_{\mathbf{k}\sigma}$ and $v_{\sigma,-}(\mathbf{k})d_{\mathbf{k}\sigma}$, respectively. The band-projected single-body Hamiltonian is then:
\begin{align}
\bar H_0
&=
\sum_{\mathbf{k} \in \text{RBZ},\sigma}
\lambda_{\mathbf{k}\sigma,-}
d^\dagger_{\mathbf{k}\sigma}d_{\mathbf{k}\sigma}
\end{align}

and the nearest-neighbor interaction of Eq.~\eqref{eq:Hint} takes the momentum-space form
\begin{align}
H_{\mathrm{int}}
&=
\frac{1}{2N_s}
\sum_{\mathbf{q} \in \text{BZ}}
\sum_{\sigma,\sigma'}
U_{\sigma\sigma'}(\mathbf{q})
:\rho_\sigma(\mathbf{q})\rho_{\sigma'}(-\mathbf{q}): ,
\end{align}
where $\rho_\sigma(\mathbf{q})
=\sum_{\mathbf p \in \text{BZ}}
c^\dagger_{\mathbf p+\mathbf{q},\sigma}
c_{\mathbf p\sigma}$ and the interaction factors are given by
\begin{align}
U_{\uparrow\uparrow}(\mathbf{q})=
U_{\downarrow\downarrow}(\mathbf{q})
&=2V(\cos q_x+\cos q_y)
\end{align}
and
\begin{align}
U_{\uparrow\downarrow}(\mathbf{q})=
U_{\downarrow\uparrow}(\mathbf{q})
&=2W(\cos q_x+\cos q_y)
\end{align}

If $|\mathbf{k},\sigma\rangle=d^\dagger_{\mathbf{k}\sigma}|0\rangle$, the projected interaction matrix element in Eq.~\eqref{eq:projected_ham} can be written as
\begin{align}
\mathcal M_{\sigma\sigma'}
(\mathbf{k}_1,\mathbf{k}_2;\mathbf{k}_3,\mathbf{k}_4)=
\sum_{\mathbf{q}\in\text{BZ}}
&U_{\sigma\sigma'}(\mathbf{q}) 
\langle\mathbf{k}_1,\sigma|
\rho_\sigma(\mathbf{q})
|\mathbf{k}_3,\sigma\rangle \nonumber\\
&\times \langle\mathbf{k}_2,\sigma'|
\rho_{\sigma'}(-\mathbf{q})
|\mathbf{k}_4,\sigma'\rangle
\end{align}

In a fixed $(N_\uparrow,N_\downarrow)$ sector, the basis consists of all Slater determinants formed by occupying $N_\uparrow$ states in the spin-up band and $N_\downarrow$ states in the spin-down band. The Hilbert space dimension (before momentum-resolution) is
\begin{align}
\dim\mathcal H(N_\uparrow,N_\downarrow)
&=\binom{N}{N_\uparrow}
\binom{N}{N_\downarrow}
\end{align}
Translation symmetry further block-diagonalizes the Hamiltonian according to the total crystal momentum, obtained by summing the occupied single-particle momenta (modulo a reciprocal lattice vector). We label these many-body momentum sectors by $(K_x,K_y)$, with $K_\alpha=0,\ldots,N_\alpha-1$, or a linearized version thereof, $\mathbf{K}=K_x+N_x K_y$.


For a $\nu=1/3$ calculation on a $3\times5$ grid, the dimension before block diagonalization by $\mathbf K$ is
\begin{align}
\dim\mathcal H(5,5)=\binom{15}{5}^2
&=9{,}018{,}009
\end{align}

The many-body spectra are obtained by sparse Lanczos diagonalization independently in each fixed $(N_\uparrow,N_\downarrow,\mathbf K)$ sector. The Hamiltonian is applied directly to Fock states, so the full many-body matrix need not be stored. 


Flux insertion is implemented through twisted boundary conditions (TBCs), $c_{\mathbf r+N_\alpha,\sigma}=e^{i\theta_\alpha^\sigma}c_{\mathbf r\sigma}$ (where $\alpha=x,y$). The allowed momenta are phase-shifted:
\begin{align}
\mathbf{k}_{\mathbf n\sigma}(\boldsymbol\theta)
&=\sum_{\alpha=x,y}
\frac{n_\alpha+\theta_\alpha^\sigma/(2\pi)}{N_\alpha}
\mathbf b_\alpha
\end{align}
which are what we use to evaluate the Bloch spinors, single-particle energies, and projected interaction matrix elements. A spin-independent flux threading satisfies $\theta_\alpha^\uparrow=\theta_\alpha^\downarrow$, whereas a spin-dependent one satisfies $\theta_\alpha^\uparrow=-\theta_\alpha^\downarrow$.

\section{Extended data}
\subsection{$\nu=\frac{1}{7}$}
\begin{figure}[!h]
\centering
\includegraphics[width=\columnwidth]{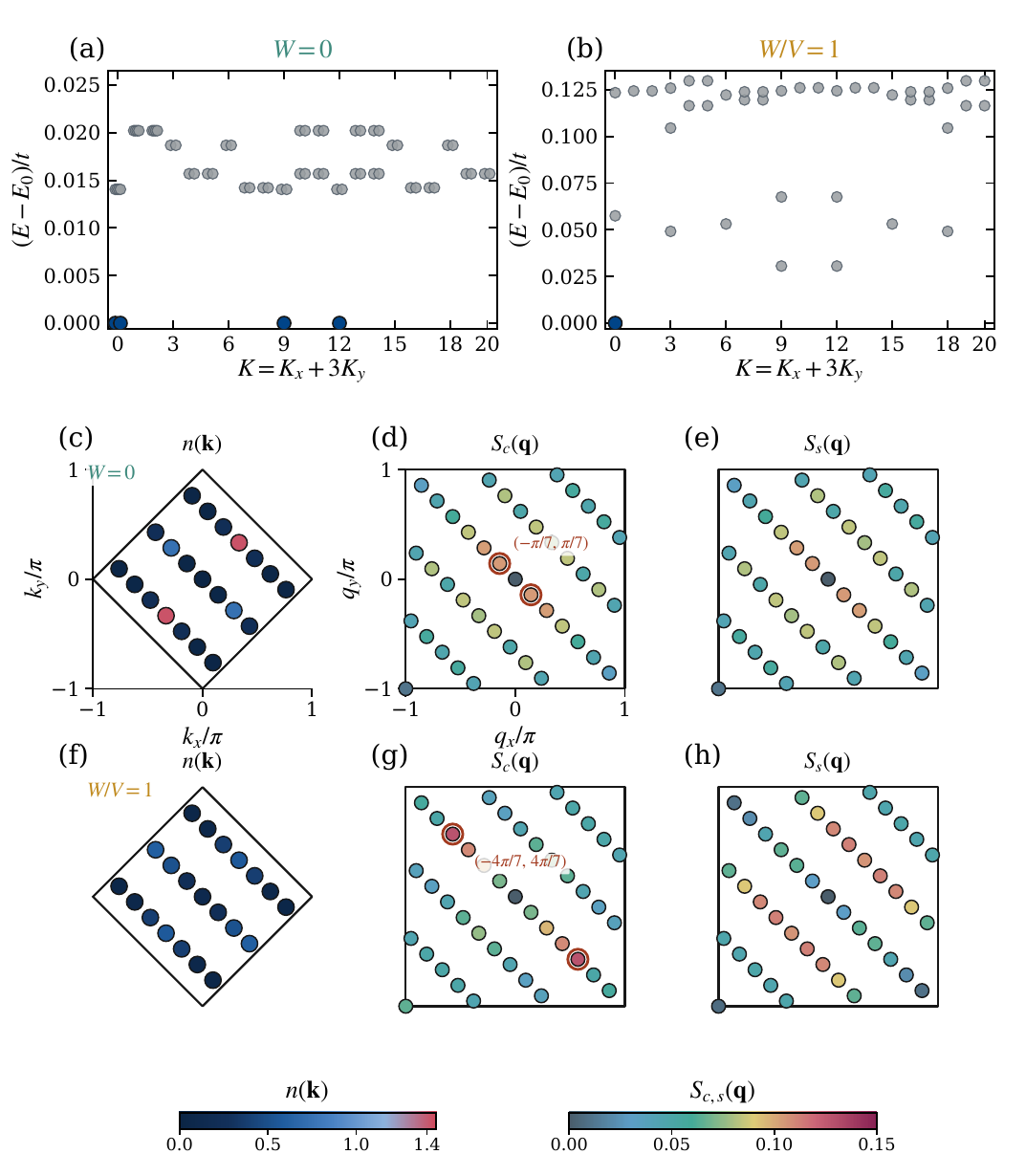}
\caption{\textbf{Spectra and correlations at $\nu=1/7$.} Exact-diagonalization results for $N_e=6$ particles with $(N_\uparrow,N_\downarrow)=(3,3)$ on a $3\times7$ torus. The physical band parameters of Sec.~\ref{sec:nu_1_3} are used with $V=1.5t$; no flat-band approximation is made. Panels (a) and (b) show the many-body levels from all momentum sectors at $W=0$ and $W/V=1$, respectively, measured from the corresponding ground-state energy and plotted against $K=K_x+3K_y$. Blue markers identify the exact finite-size ground states. Panels (c)--(e) are equal-weight averages over the four degenerate $W=0$ ground states, whereas panels (f)--(h) use the unique ground state at $W/V=1$. The panels show the spin-summed projected-band occupation $n(\mathbf{k})$ and the connected microscopic charge and longitudinal-spin structure factors $S_c(\mathbf{q})$ and $S_s(\mathbf{q})$. The highlighted circles mark the dominant nonzero-momentum charge correlations at $\mathbf{q}=(-\pi/7,\pi/7)$ for $W=0$ and $\mathbf{q}=(-4\pi/7,4\pi/7)$ for $W/V=1$.}
\label{fig:nu_1_7_spectra_correlations}
\end{figure}

Figure~\ref{fig:nu_1_7_spectra_correlations} shows the spectrum, momentum occupation, and density correlations for the dispersive $3\times7$ torus discussed in the main text.

\subsection{$\nu=\frac{1}{2}$}
\subsubsection{$W=0$}
\begin{figure*}[!t]
\centering
\includegraphics[width=0.8\textwidth]{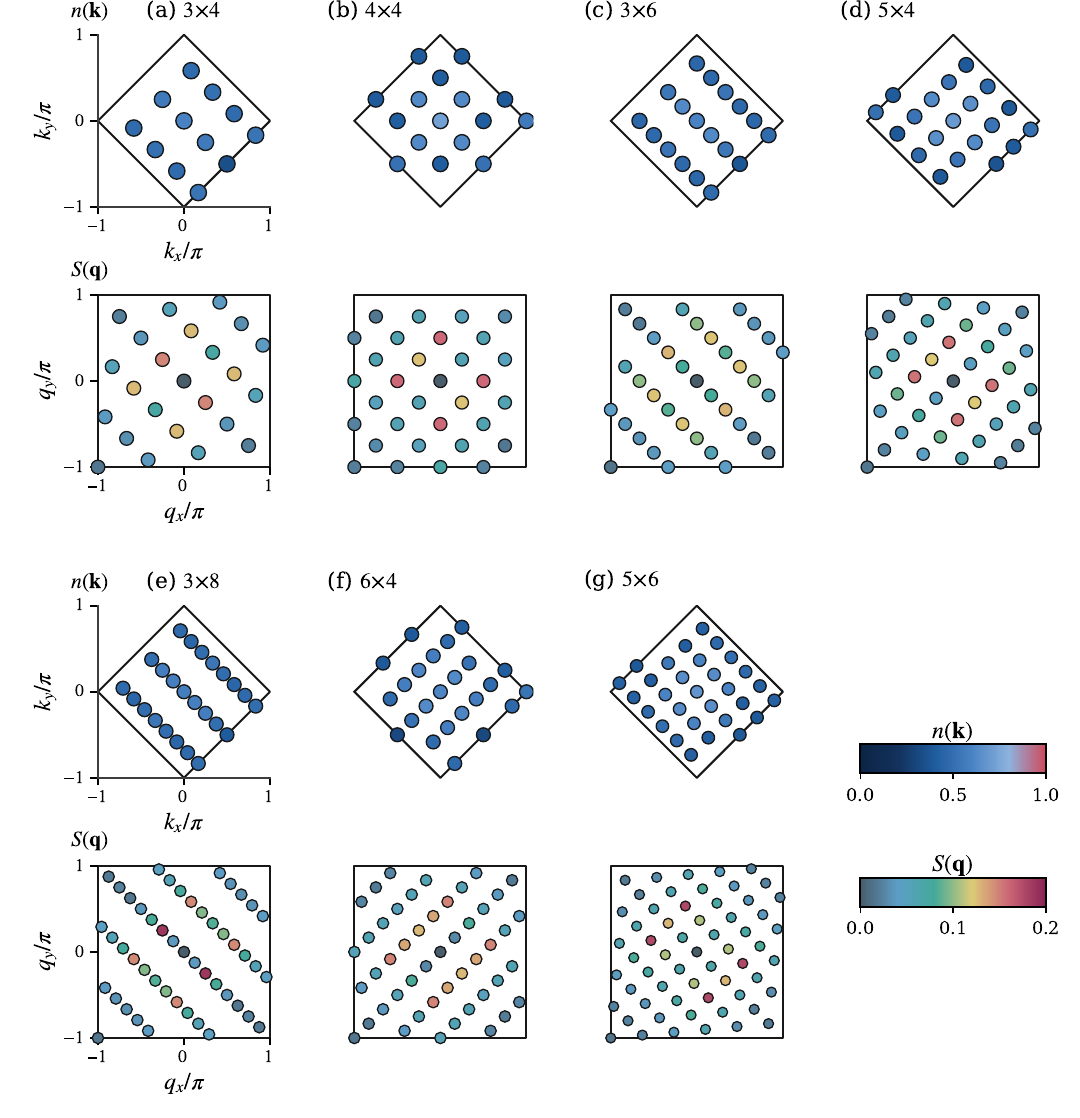}
\caption{\textbf{Momentum occupation and density correlations at $\nu=1/2$.} Exact-diagonalization results for one fully polarized lower stDDW band at half filling on (a) $3\times4$, (b) $4\times4$, (c) $3\times6$, (d) $5\times4$, (e) $3\times8$, (f) $6\times4$, and (g) $5\times6$ tori. The band parameters of Sec.~\ref{sec:nu_1_3} are used, with $V=1.5t$ and $W=0$. The upper plots show $n(\mathbf{k})$ on the RBZ. The lower plots show $S(\mathbf{q})$ on the full BZ.}
\label{fig:nu_1_2_correlations_all_geoms}
\end{figure*}

\paragraph{Static correlations.}
To test for charge order, Fig.~\ref{fig:nu_1_2_correlations_all_geoms} shows the momentum occupation and connected band-projected density structure factor. At PBCs, both observables are averaged over the complete degenerate ground state space.

The occupation remains geometry dependent, while the largest nonzero momentum structure factor stays between $0.129$ and $0.187$ without selecting a common wave vector or growing with system area. Thus, these clusters show no evidence of density order.

\paragraph{Band-projected spectral function.}

To examine the charged excitations of the half-filled state, we calculate the zero-temperature addition and removal spectrum of a spin-polarized lower stDDW band. Recall the definition of $d_{\mathbf{k}\sigma}$ in Eq.~\eqref{eq:projected_band_operator}. The calculation is restricted to the projected band and excludes the additional orbital matrix elements associated with a particular tunneling probe. It is therefore not directly the spectral function that would be measured by ARPES.


We use the parameters of Sec.~\ref{sec:nu_1_3}. The interacting calculations use $V=1.5t$ and $W=0$, while $V=0$ is the noninteracting reference. We study $3\times4$, $4\times4$, $3\times6$, $5\times4$, $3\times8$, and $6\times4$ tori, with $N_\sigma=N_xN_y/2$ particles. The $3\times8$ and $6\times4$ clusters have the same number of orbitals but different aspect ratios.

\begin{figure*}[p]
\centering
\includegraphics[width=0.72\textwidth]{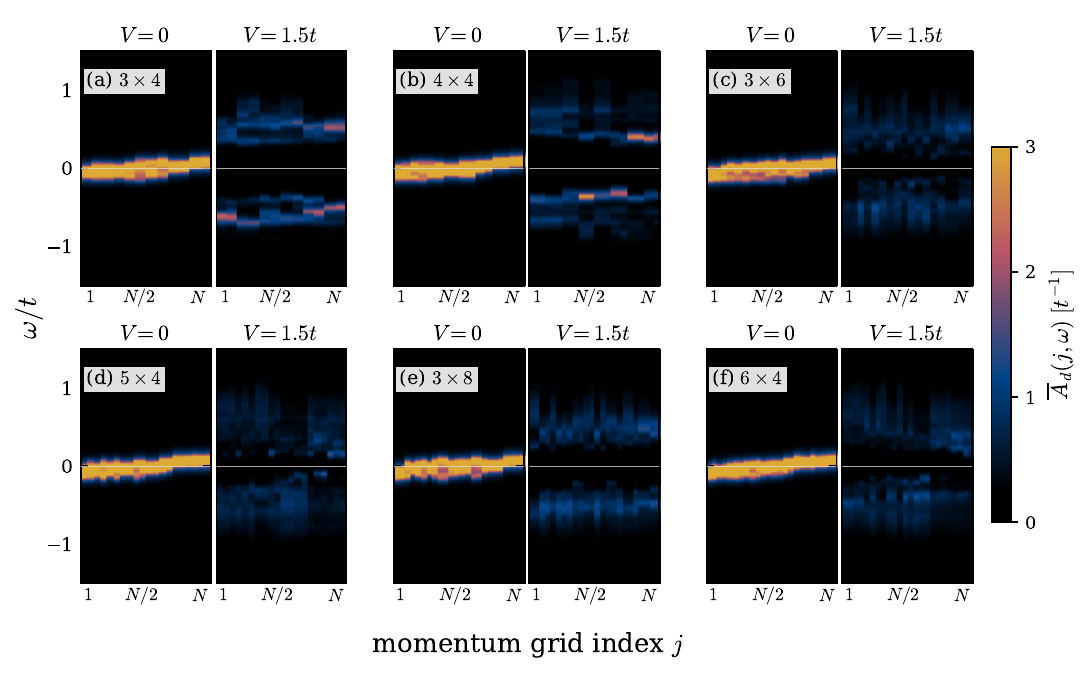}
\caption{\textbf{Band-projected spectra at $\nu=1/2$.} Addition and removal spectra for one spin-resolved lower stDDW band on (a) $3\times4$, (b) $4\times4$, (c) $3\times6$, (d) $5\times4$, (e) $3\times8$, and (f) $6\times4$ tori. For each geometry, the left panel shows the noninteracting  $V=0$ reference and the right panel shows $V=1.5t$, with $W=0$. The spectra are averaged over eight twist angles according to Eq.~\eqref{eq:nu_1_2_spectral_twist_average}. The momentum-grid index $j=1,\ldots,N_xN_y$ orders the momenta by their twist-averaged noninteracting energy. Each pole is broadened by $\eta=0.04t$. Interactions redistribute the sharp noninteracting spectral weight over a broad range of energies.}
\label{fig:nu_1_2_spectral_fun}
\medskip
\includegraphics[width=0.60\textwidth]{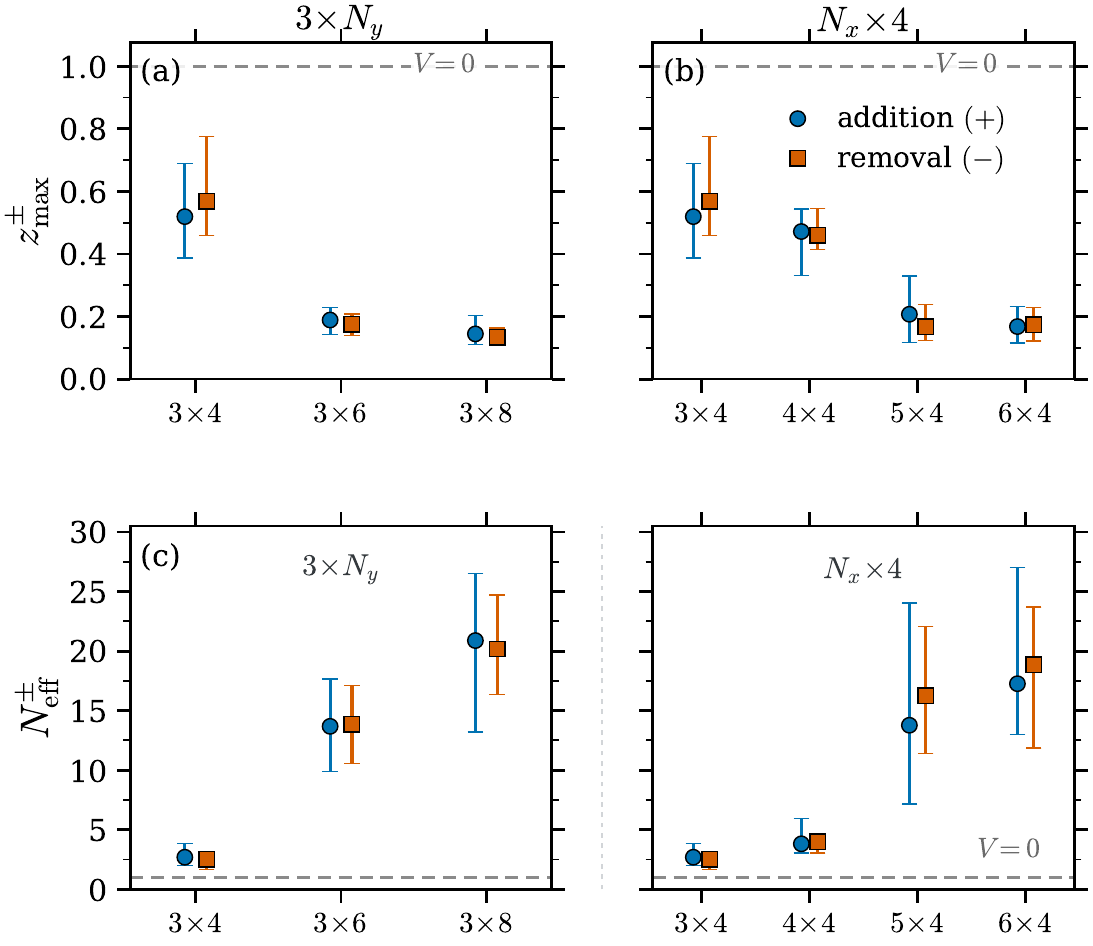}
\caption{\textbf{Size dependence of the projected-electron spectral weight.} The largest normalized pole weight $z_{\max}^{\pm}$ [panels (a) and (b)] and the pole weight participation ratio $N_{\mathrm{eff}}^\pm$ [panels (c) and (d)] are shown for the $3\times N_y$ and $N_x\times4$ sequences at $V=1.5t$ and $W=0$. Circles and squares represent electron addition and removal, respectively. The center values are medians over all momenta and eight TBCs, while the bars show the corresponding interquartile ranges (not to be confused with uncertainties). The dashed lines denote the exact noninteracting values, $z_{\max}^\pm=N_{\mathrm{eff}}^\pm=1$. The decreasing $z_{\max}^\pm$ and increasing $N_{\mathrm{eff}}^\pm$ show that the band-projected weight gets distributed over progressively more poles on the cluster geometries computed here.}
\label{fig:nu_1_2_spectral_weights}
\end{figure*}

To reduce sensitivity to the discrete momentum grid, each geometry is evaluated at the same eight generic twist angles. These twist angles are meant to sample different placements of the discrete momentum grid and should not be interpreted as a quadrature over all boundary conditions. For each twist, we first determine the global $N_\sigma$-particle ground state and the lowest energies in the $N_\sigma\pm1$ sectors.


For a fixed twist angle, addition and removal energies are measured relative to
\begin{align}
\mu&=
\frac{E_0^{N_\sigma+1}-E_0^{N_\sigma-1}}{2}
\end{align}
Suppressing the spin and twist indices, we define
\begin{align}
\omega_m^+
&=E_m^{N_\sigma{+}1}-E_0^{N_\sigma}-\mu,
\quad
\omega_m^-
=E_0^{N_\sigma}-E_m^{N_\sigma{-}1}-\mu,
\\
Z_m^+(\mathbf{k})
&=\bigl|\langle m,\!N_\sigma{+}1|d_{\mathbf{k}}^{\dagger}|0,\!N_\sigma\rangle\bigr|^2,
\nonumber\\
Z_m^-(\mathbf{k})
&=\bigl|\langle m,\!N_\sigma{-}1|d_{\mathbf{k}}|0,\!N_\sigma\rangle\bigr|^2
\end{align}
The spectral function is
\begin{align}
A_d^\pm(\mathbf{k},\omega)
&=\sum_m Z_m^\pm(\mathbf{k})L_\eta(\omega-\omega_m^\pm),
\nonumber\\
A_d(\mathbf{k},\omega)
&=A_d^+(\mathbf{k},\omega)+A_d^-(\mathbf{k},\omega)
\end{align}
where $L_\eta(x)=\eta/[\pi(x^2+\eta^2)]$. We use $\eta=0.04t$ for visualization. The exact finite-size spectrum consists of discrete poles, so $\eta$ is an artificial broadening rather than a physical decay rate. With the convention used above, the lowest addition and removal energies occur at $\pm\Delta_c/2$, where $\Delta_c=
E_0^{N_\sigma+1}+E_0^{N_\sigma-1}-2E_0^{N_\sigma}$.
If the ground state momentum is $\mathbf K_0$, addition and removal at momentum $\mathbf{k}$ access the sectors $\mathbf K_0+\mathbf{k}$ and $\mathbf K_0-\mathbf{k}$, respectively.


We begin Lanczos iteration from:
\begin{align}
|f_{\mathbf{k}}^+\rangle&=d_{\mathbf{k}}^{\dagger}|0,N_\sigma\rangle,
\qquad
|f_{\mathbf{k}}^-\rangle=d_{\mathbf{k}}|0,N_\sigma\rangle
\end{align}
whose norms give:
\begin{align}
\sum_m Z_m^+(\mathbf{k})&=1-n(\mathbf{k}),
\qquad
\sum_m Z_m^-(\mathbf{k})=n(\mathbf{k}),
\label{eq:nu_1_2_spectral_sum_rules}
\end{align}
where $n(\mathbf{k})=\langle d_{\mathbf{k}}^{\dagger}d_{\mathbf{k}}\rangle$. We use a fully reorthogonalized Lanczos basis of at most $L=400$ vectors and ensure that the relations in Eq.~\eqref{eq:nu_1_2_spectral_sum_rules} are satisfied. When the Krylov space spans the target momentum sector, its poles are exact eigenstates. Otherwise, the recursion should be interpreted as giving a discrete representation of the spectral measure. For an example benchmark, comparing the $L=400$ calculation against the complete
$L=715$ calculation for the 4$\times$4 torus gives differences below $10^{-14}$
in the normalized broadened spectrum and $10^{-5}$ in the
participation ratio defined below. The ``sum rules'' of
Eq.~\eqref{eq:nu_1_2_spectral_sum_rules} are satisfied to within
$10^{-14}$ at every sampled momentum and twist angle.


In the plots below, we average the broadened spectra over eight twist angles
\begin{align}
\overline A_d(\mathbf n,\omega)
&=
\frac{1}{8}\sum_{a=1}^{8}
A_d\!\left(\mathbf{k}_{\mathbf n}^{(a)},\omega;\boldsymbol\theta_a\right)
\label{eq:nu_1_2_spectral_twist_average}
\end{align}
Here $\mathbf n$ labels the same discrete grid point at each twist, while $\mathbf{k}_{\mathbf n}^{(a)}$ is its shifted momentum. Each spectrum is aligned using its own midpoint chemical potential before averaging.


In Fig.~\ref{fig:nu_1_2_spectral_fun}, the grid points are ordered by their twist-averaged noninteracting band energy. The  $j$ index labels an ordered list of momenta and should not be interpreted as a continuous momentum path. At $V=0$, each allowed addition or removal process creates a single exact eigenstate. The corresponding spectrum therefore contains one pole carrying all of its weight. At $V=1.5t$, the weight is distributed over many energies on every cluster. Eq.~\eqref{eq:nu_1_2_spectral_twist_average} is used only for this visualization. All quantitative measures below are evaluated separately at each momentum and twist before averaging.


The total addition and removal weights depend on $n(\mathbf{k})$. For each momentum and twist angle, we denote the pole weights normalized by their respective totals by $\widetilde Z_m^\pm$ and define:
\begin{align}
z_{\max}^{\pm}&=\max_m\widetilde Z_m^\pm, \qquad
N_{\mathrm{eff}}^{\pm}=\left[\sum_m\left(\widetilde Z_m^\pm\right)^2\right]^{-1}
\end{align}
where the first quantity is the fraction of the addition or removal weight carried by its largest pole. The second is its participation ratio. One should be careful to avoid calling these quantities a thermodynamic ``quasiparticle residue'' since they are evaluated over finite-system momenta. 
Figure~\ref{fig:nu_1_2_spectral_weights} summarizes the median and interquartile range of these quantities over all momenta and twist angles. At $V=0$, every allowed addition or removal spectrum contains one pole, so $z_{\max}^\pm=N_{\mathrm{eff}}^\pm=1$. At $V=1.5t$, the median $z_{\max}^\pm$ decreases from approximately $0.5$ on $3\times4$ to $0.13$--$0.17$ on the largest clusters. Over the same sizes, $N_{\mathrm{eff}}^\pm$ increases from approximately $2.5$ to $17$--$21$. The trend is qualitatively similar for both addition and removal and along both finite-size sequences.

\paragraph{Comparison with the lowest Landau level.}

\begin{figure*}[p]
\centering
\includegraphics[width=0.68\textwidth]{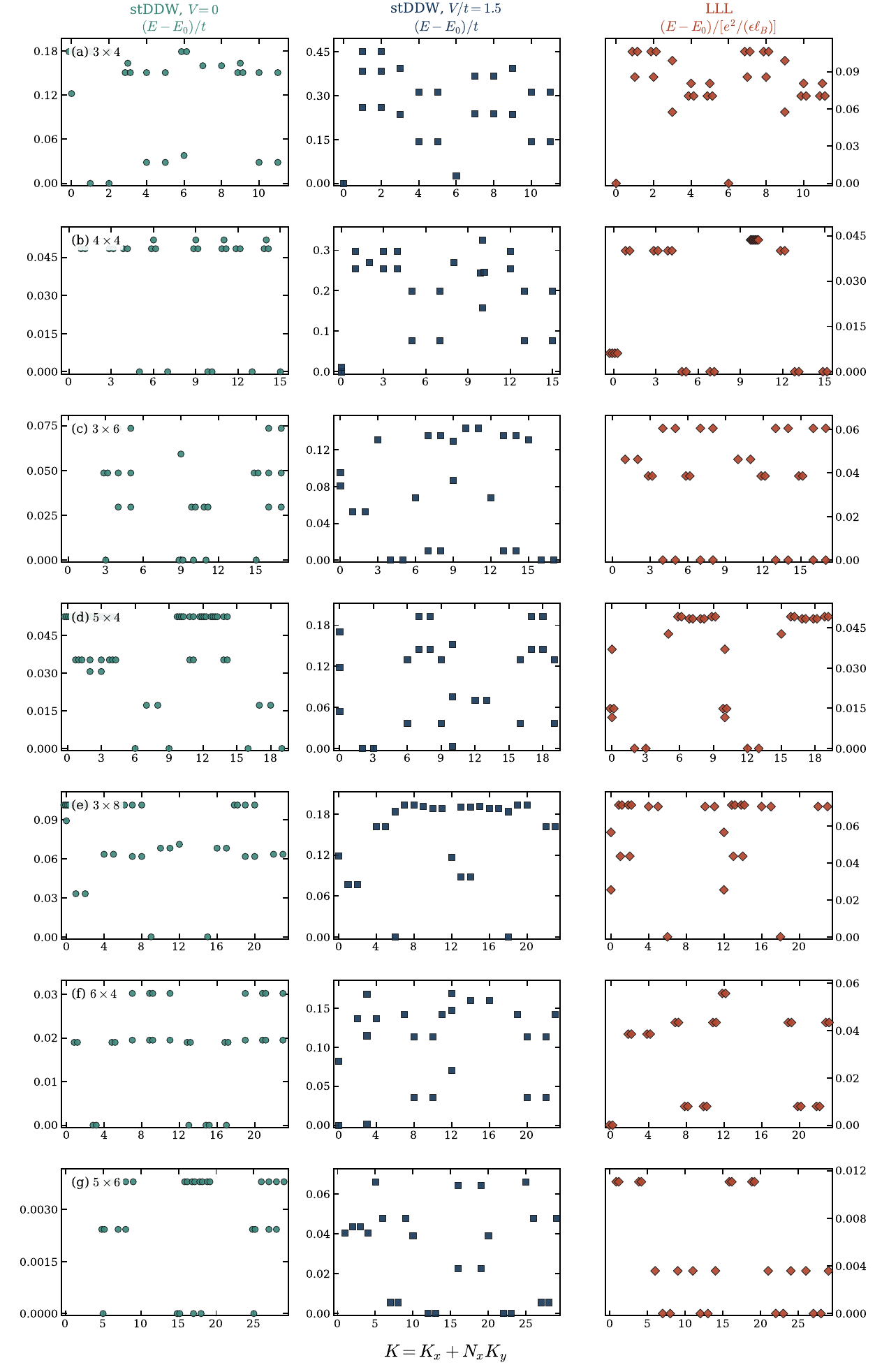}
\caption{\textbf{Comparison of the noninteracting stDDW, interacting stDDW, and LLL spectra at half filling.} Many-body spectra for a single-component Hamiltonian on (a) $3\times4$, (b) $4\times4$, (c) $3\times6$, (d) $5\times4$, (e) $3\times8$, (f) $6\times4$, and (g) $5\times6$ tori. Each system contains $N=N_xN_y$ orbitals and $N_\sigma=N/2$ particles. Green circles show the exact $V=0$ many-body spectrum obtained from the dispersive lower stDDW band, and blue squares show the spectrum at $V=1.5t$ and $W=0$. Their excitation energies are given in units of $t$. Orange diamonds show the Coulomb LLL spectrum at the same particle number and aspect ratio, in units of $e^2/(\epsilon\ell_B)$. Every spectrum is measured from its own ground state energy. The lowest 24 states are shown, together with any exact multiplet crossing this boundary. The twofold LLL center of mass degeneracy is restored before the continuum momenta are folded into the lattice Brillouin zone according to Eq.~\eqref{eq:nu_1_2_lll_folding}. }
\label{fig:nu_1_2_lll_comparison}
\end{figure*}

We compare the spin-polarized stDDW spectrum with the Coulomb spectrum of a half-filled lowest Landau level (LLL), following Refs.~\cite{Dong2023,Goldman2023}, which used finite size spectral comparisons to identify anomalous composite Fermi liquids in even-denominator filled Chern bands. Here, the LLL is simply used as a reference and should not be interpreted as an effective Hamiltonian derived from the stDDW model. As discussed in Ref.~\cite{Wang2021}, even for an ``ideal'' Chern band, mapping a projected lattice interaction into the LLL generally produces center of mass dependent interaction terms rather than the ordinary Coulomb Hamiltonian. We therefore look at the momentum sectors, degeneracies, and ordering of the lowest levels, but not their absolute energies or the overlap between their many-body wave functions since that is not straightforward to define.


For an $N_x\times N_y$ geometry, the continuum LLL contains $N_\phi=N_xN_y$ flux quanta and $N_\sigma=N_\phi/2$ particles. Its area is $A=2\pi N_\phi\ell_B^2$, and its aspect ratio $L_x/L_y=N_x/N_y$ matches the lattice geometry. Since the LLL kinetic energy is constant, the reference Hamiltonian contains only the projected Coulomb interaction,
\begin{align}
H_{\mathrm{LLL}}
&=
\frac{1}{2A}
\sum_{\mathbf{q}\neq0}
\frac{2\pi e^2}{\epsilon |\mathbf{q}|}
e^{-|\mathbf{q}|^2\ell_B^2/2}
:\overline{\rho}(-\mathbf{q})\overline{\rho}(\mathbf{q}):\,
\end{align}
Here $\overline{\rho}(\mathbf{q})$ is the guiding-center density and the neutralizing background removes the $\mathbf{q}=0$ term. The LLL energies are in units of $e^2/(\epsilon\ell_B)$.


The lattice and LLL calculations have different translation algebras, so their momentum labels must be related by magnetic translation symmetry before the spectra can be compared \cite{Bernevig2012}. At $N_\sigma/N_\phi=1/2$, magnetic translation symmetry gives an exact twofold center of mass degeneracy \cite{Haldane1985}. Our LLL calculations are reduced by translation symmetry and retain one representative from each pair. We first restore the omitted partner and then fold both states into the $N_x\times N_y$ lattice Brillouin zone. In our momentum convention,
\begin{align}
K_x&=\kappa_x \bmod N_x,
\nonumber\\
K_y&=(\kappa_y+sN_\sigma)\bmod N_y,
\qquad s=0,1
\label{eq:nu_1_2_lll_folding}
\end{align}
for a $C=1$ spin sector, where $(\kappa_x,\kappa_y)$ label the folded LLL sectors. Restoring and folding the LLL spectrum reproduces the full Hilbert-space dimension. Figure~\ref{fig:nu_1_2_lll_comparison} compares the noninteracting stDDW spectrum, the interacting stDDW spectrum at $V=1.5t$, and the Coulomb LLL spectrum for each geometry. The agreement is strongest on the smallest torus. On larger clusters, several low-lying LLL momentum sectors and multiplicities remain visible in the interacting stDDW spectrum, whereas the noninteracting spectrum follows a different finite-size filling pattern. The correspondence is nevertheless incomplete, since additional stDDW states intervene and the sequence of LLL levels is not reproduced exactly for every geometry.

\subsubsection{$W\neq0$}
\begin{figure}[!t]
\centering
\includegraphics[width=\columnwidth]{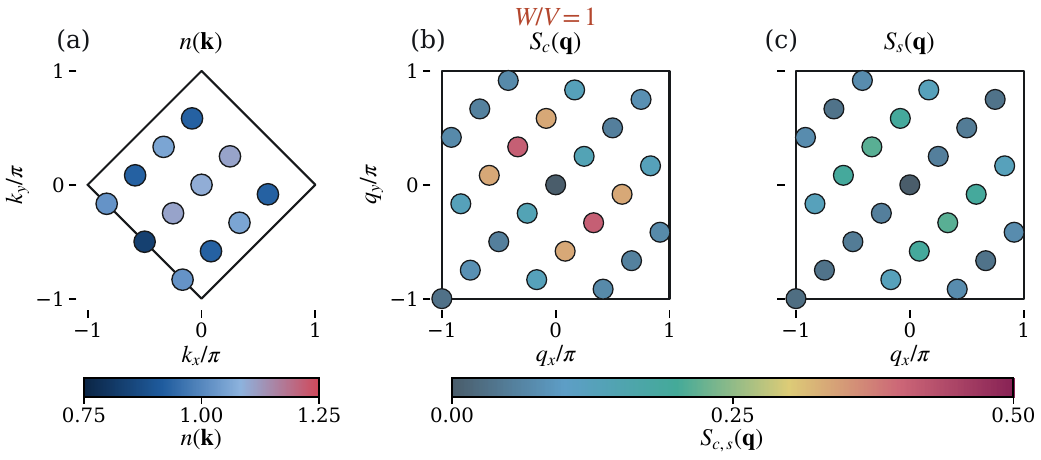}
\caption{\textbf{Momentum occupation and structure factor of the spin-coupled half-filled state.} Momentum occupation (a), charge structure factor (b), and spin structure factor (c) for $N_e=12$ particles on the $4\times3$ torus at $V=W=1.5t$ ($W/V=1$). All observables are equal-weight averages over the lowest two states in momentum sectors $(K_x,K_y)=(0,0)$ and $(2,0)$. The data show modest variation of $n(\mathbf{k})$ and a feature at $\mathbf{q}=(-\pi/3,\pi/3)$ in the charge structure factor.}
\label{fig:nu_1_2_spinful_correlations}
\end{figure}

\begin{figure}[!t]
\centering
\includegraphics[width=\columnwidth]{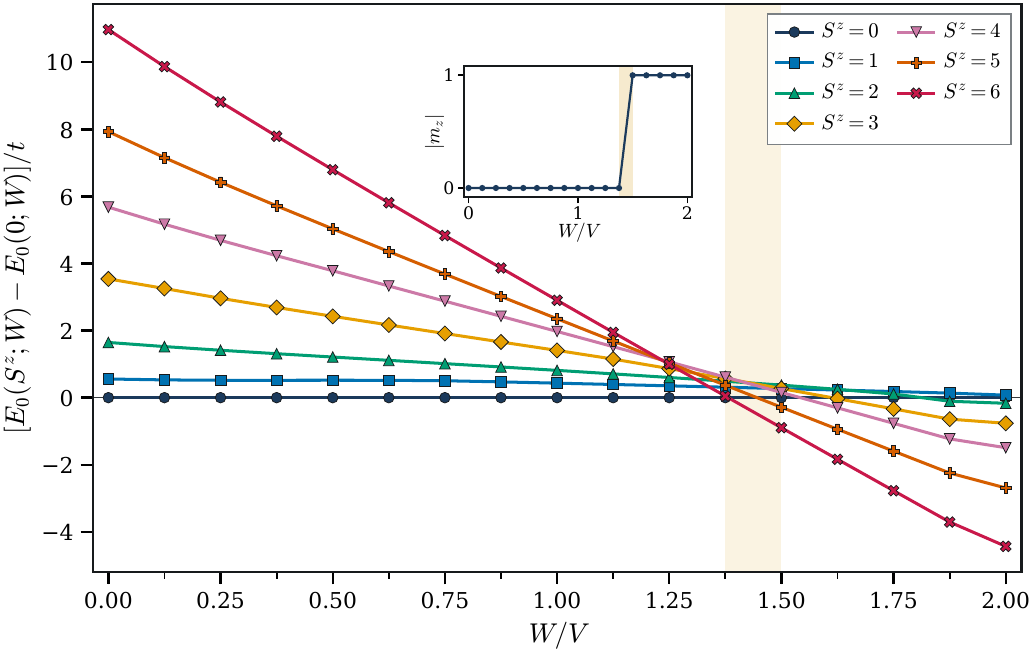}
\caption{\textbf{Spin-polarization transition at half filling.} Ground state energies in all conserved spin sectors for $N_e=12$ particles on the $4\times3$ torus, with $V=1.5t$ and the single-particle parameters stated in the main text. For each $S^z=0,1,\ldots,6$, the lowest energy is minimized over every many-body momentum sector and plotted relative to the $S_z=0$ energy at the same value of $W$. The fully polarized sector crosses directly below the unpolarized branch between the sampled values $W/V=1.375$ and $1.5$, while no partially polarized sector becomes the absolute ground state. The inset shows the absolute value of ground state polarization $|m_z|=|N_\uparrow-N_\downarrow|/N_e$. Its jump from $0$ to $1$ is consistent with a first order transition into quantum anomalous Hall ferromagnets.}
\label{fig:nu_1_2_spin_transition}
\end{figure}

Figures~\ref{fig:nu_1_2_spinful_correlations} and \ref{fig:nu_1_2_spin_transition} provide additional data for the spinful case on a $4\times3$ torus. The first shows the static observables at $W/V=1$, while the second tracks the lowest energy in each conserved-spin sector as $W/V$ is increased and shows the transition to full spin polarization.

\clearpage

\end{document}